\documentclass[11pt]{article}
\usepackage{amsmath,amssymb,mathtools}
\usepackage[margin=1in]{geometry}
\usepackage[numbers,compress]{natbib}
\usepackage{hyperref}
\usepackage{subcaption}
\usepackage{braket}
\usepackage{xcolor}
\usepackage{cleveref}
\usepackage{graphicx}
\usepackage{bm}
\usepackage{doi}
\usepackage{url}
\usepackage{authblk}
\usepackage[normalem]{ulem}

\newcommand{\sgn}{\operatorname{sgn}}
\renewcommand{\d}{{\rm d}}

\title{A dynamical systems perspective on the thermodynamics of late-time cosmology}

\date{}

\author[1]{Dipayan Mukherjee\thanks{dipayanmkh@gmail.com}}
\affil[1]{\small \emph{Raman Research Institute, C.~V.~Raman Avenue, Sadashivanagar, Bengaluru 560080, India.}}

\author[2,3]{Harkirat Singh Sahota\thanks{harkirat221@gmail.com}}
\affil[2]{\small \emph{Department of Physics,  Indian Institute of Technology Delhi, Hauz Khas, New Delhi, 110016, India.}}
\affil[3]{Department of Physics, Indian Institute of Technology Madras, Chennai 600036, India.}

\author[4,5]{Swati Gavas\thanks{swatigavas47@gmail.com}}
\affil[4]{\small \emph{School of Physical Sciences, National Institute of Science Education and Research, Jatni 752050, India.}}
\affil[5]{\small \emph{Homi Bhabha National Institute, Training School Complex, Anushaktinagar, Mumbai 400094, India.}}

\begin{document}
\maketitle
\begin{abstract}
A thermodynamic description of cosmological spacetimes may provide insights into the fundamentals of the cosmic evolution that remain otherwise obscure, similar to `black hole thermodynamics'. We investigate the thermodynamic properties of late-time cosmological evolution using the dynamical systems approach, focusing on $\Lambda${}CDM model and scalar field models with exponential potentials. Thermodynamic quantities obtained through the Hayward-Kodama formalism are mapped onto the phase-space of these models. 
Specifically, we express the thermodynamic quantities as functions of the phase-space variables, allowing us to study the thermodynamic behavior across the phase space, particularly at the critical points. We focus on thermodynamic stability and phase transitions, analyzed in an initial condition-independent manner. 
In these models, the universe inevitably undergoes a thermodynamic phase transition, marked by diverging specific heats, irrespective of its initial configuration. We further demonstrate that the thermodynamic stability can occur only during an accelerating phase of the universe. For $\Lambda$CDM and quintessence models, the necessary stability conditions are never satisfied anywhere in the phase space, rendering both models thermodynamically unstable within the Hayward-Kodama framework
and the canonical ensemble based stability criteria.
Interestingly, the phantom models, although dynamically unstable, allow for the universe to attain thermodynamic stability in its asymptotic future.
This can indicate the limitations of applying canonical ensemble based thermodynamic stability criteria to cosmological horizons.
Through these archetypal descriptions of late-time cosmology, we show that the dynamical system approach is a robust framework to probe the thermodynamic aspects of cosmological evolution. 
\end{abstract}

\section{Introduction}

A thermodynamic perspective of spacetime geometry originated from Bekenstein's insight into the entropy of black hole geometry~\cite{Bekenstein1973,Bekenstein:1974ax} and with the subsequent formulation of the `laws of black hole mechanics' by Bardeen, Carter, and Hawking (BCH)~\cite{Bardeen1973}. This formal analogy of black hole geometry with a thermodynamical system holds when the surface gravity at the event horizon is identified with temperature and area of the horizon with entropy, enabling the investigation of thermodynamical aspects of this gravitational system \cite{Davies:1977bgr,Candelas:1977zz,Davies:1978zz,Sokolowski:1980uva,Davies:1989ey,Padmanabhan:1989gm,Pavon:1991kh,Katz:1993up,Kaburaki:1993ah,Shen:2005nu,Kothawala:2007em,Banerjee:2010da,Banerjee:2011cz,Thomas:2012zzc,Czinner:2015eyk,Majhi:2016txt,Banerjee:2016nse,Bhattacharya:2019awq,Bargueno:2024mys,Shahzad2024-qz}. Hawking's discovery of particle creation in black hole geometry~\cite{Hawking:1974rv,Hawking1975} lent further credence to this geometric–thermodynamic correspondence, which nowadays is understood to be of quantum origin \cite{Wald:1999vt,Carlip:2014pma,Mohd:2023hhu}. This correspondence, in fact, extends beyond black holes and is valid more generally for spacetime geometries featuring horizons \cite{Padmanabhan:2002ha,Padmanabhan:2009vy, Padmanabhan:2022jjt}, e.g., cosmological horizon in de Sitter spacetime \cite{Gibbons:1977mu} or Rindler horizon in Minkowski spacetime~\cite{Unruh:1976db}. 

It is conceivable that the thermodynamic properties of cosmological spacetimes will also reveal fundamental features of the cosmic evolution that are otherwise unapparent. However, unlike the case of black hole thermodynamics, the understanding of a thermodynamic description of cosmological spacetimes is still lacking and remains an area of active research~\cite{faraoni2013scalar}. For a typical cosmological spacetime, barring the case of a de Sitter universe, the horizon is time-evolving. Therefore, the standard considerations concerning the stationary horizon need to be generalized using the notion of an apparent horizon. In this regard, the Hayward-Kodama formalism~\cite{Kodama:1979vn, Hayward:1997jp, Hayward:2008jq} offers a well-motivated starting point, allowing one to investigate the thermodynamic aspects of the universe such as the laws of thermodynamics~\cite{Cai:2005ra,Izquierdo:2005ku} and thermodynamical stability of cosmological evolution~\cite{Luciano:2022ffn, Duary:2019dfu,Duary:2023nnf,Chakrabarti:2024amb, Saha:2024mwn}. In particular, it is anticipated that thermodynamical aspects based on this prescription may shed some light on the mysteries of dark energy~\cite{Luciano:2022ffn, Duary:2019dfu,Duary:2023nnf,Chakrabarti:2024amb, Saha:2024mwn, Luongo:2012dv, Setare:2006vz, Gong:2006ma, Barboza:2015bha, Bhandari:2017cow,Bahamonde2018-qw,Chetry:2019tqd, Kong:2021dqd, Kong:2022xny, Abdusattar:2021wfv, Abdusattar:2023hlj, banihashemi2023minus,Sebastiani2023-sr,Cruz2023-sf,Panda2024-dg,Singh2024-wq,Cruz2024-iq, Chakrabarti:2025qsv,Samaddar_2025,Caro:2025}.

Recently, the questions of thermodynamic stability and phase transitions in different cosmological models have been addressed in the Hayward-Kodama formalism, focusing on whether these thermodynamic phenomena are fundamentally related to the cosmological quantities~\cite{Duary:2019dfu, Duary:2023nnf, Chakrabarti:2024amb,Saha:2024mwn}. For example, it was speculated in~\cite{Duary:2023nnf} that the transition of the universe's expansion from a decelerating phase to an accelerating one may be associated with a second-order thermodynamic phase transition marked by diverging specific heats, hinting at a close relation between the thermodynamic and cosmological aspects. Notably,~\cite{Saha:2024mwn} points out that the ratio of specific heats at constant pressure and volume is precisely the (negative) deceleration parameter of the evolution of a flat FLRW universe, thereby further suggesting such a connection. It is further expected that thermodynamic viability could serve as a diagnostic criterion when assessing equivalent models of late-time cosmic acceleration \cite{Duary:2019dfu}. 

We should note that such thermodynamic characterizations in the FLRW universe can be elusive due to the time-dependent nature of the spacetime. Unlike that of black holes, the thermodynamic quantities associated with the universe's expansion --- and the relations between them --- evolve in time and thus depend on the initial configuration of the universe. It is crucial to examine whether a thermodynamic phenomenon, such as a phase transition, is inherently linked to the cosmological model or is an artifact of the chosen initial condition. For this reason, analyzing and inferring thermodynamic features in a manner that is independent of initial conditions is imperative, as it shall reveal whether they are generic to the cosmological model itself. 

In this light, we study the thermodynamic aspects of cosmological evolution using the dynamical systems approach, where the evolution of the universe is represented as an orbit in the phase space of relative energy contributions of different matter species (see~\cite{Bahamonde:2018pr, amendola2010dark, copeland2006dynamics, garcia2015introduction, fay2013ten,Xu:2012jf,Leon:2012mt,Leon:2013qh,Leon:2014yua} and references therein). Given a cosmological model, the associated thermodynamic quantities --- such as entropy and specific heats --- are thus mapped onto the phase space. 
Crucially, we recast these thermodynamic variables explicitly in terms of the dynamical phase space coordinates. This formulation allows us to analyze thermodynamic behavior globally across the phase space, with a particular focus on how these quantities behave at the critical points.
This approach provides two main advantages: First, because the phase space takes into account all physically viable cosmological evolutions, thermodynamic properties can be examined across the full set of orbits without the need of specifying initial conditions. Second, the dynamical system analysis reveals stable, unstable, and saddle-type critical points on the phase space: The stable and unstable points generally act as future and past attractors of the orbits, while saddle points temporarily attracts and then repel trajectories, thus describing transient phases. Starting from an arbitrary energy budget, the universe `tends to follow' orbits connecting these past to future attractors. Therefore, any thermodynamic phenomena encountered along these orbits --- such as phase transitions or regions of thermodynamic stability --- would be generic to the cosmological model, independent of the initial conditions. 
Furthermore, with thermodynamic quantities mapped onto the phase space, one can directly compare the notion of dynamical stability derived from critical point analysis with that of thermodynamic stability --- associated with maximum entropy --- and examine whether the two coincide.

In this paper, we consider the concordance ($\Lambda${}CDM) model of cosmology~\cite{dodelson2021concordance}, in which dark energy --- implemented through the cosmological constant $\Lambda$ --- dominates the late-time universe, with non-relativistic matter (dust) and radiation as subdominant components. We then extend the analysis to dynamical dark energy models, namely a quintessence model with an exponential potential~\cite{Copeland:1998PhRvD, Urena-Lopez:2012JCAP} and phantom model of non-canonical scalar field with exponential potential~\cite{Caldwell:1999ew,Caldwell:2003vq}. For all cases, we investigate thermodynamic aspects within the phase space formalism as discussed above, mainly focusing on thermodynamic stability and the occurrence of thermodynamic phase transitions. We also examine whether the transition into the accelerating phase corresponds to a thermodynamic phase transition.

This paper is organized as follows: In Sec.~\ref{sec:HK_temp} we briefly introduce the Hayward-Kodama temperature associated with an FLRW universe. We investigate the thermodynamical aspects in the phase space of the $\Lambda${}CDM, quintessence and phantom models in Sec.~\ref{sec:thermo}. We conclude with a summary and discussion in Sec.~\ref{sec:summary}.

\textit{The metric signature is taken to be $(-,+,+,+)$. Overdots denote derivatives with respect to the comoving time coordinate $t$, while primes denote derivatives with respect to the e-fold $\tau \equiv \ln a$. We use units for which $c = 8 \pi G = 1. $}

\section{Hayward-Kodama temperature and thermodynamics of FLRW universe}
\label{sec:HK_temp}
We consider a spatially flat FLRW spacetime with line element
\begin{align}
    \d s^2=-\d t^2+ a ^2(t) \delta_{ij} \d x^i \d x^j,
\end{align}
where $a$ is the scale factor. The evolution of the universe is governed by the Friedmann and Raychaudhuri equations
\begin{align}
    H^2 &= \frac{1}{3} \sum_{I} \rho_I,\\
\dot{H} &=-\frac{1}{2} \sum_{I} (\rho_I + P_I).
\end{align}
Here $H=\dot{a}/a$ is the Hubble parameter, $\rho_I$ and $P_I$ are the energy density and pressure associated with the matter species $I$. The evolution of the matter sectors are governed by the corresponding continuity equations
\begin{align}
    \dot{\rho}_I + 3 H \rho_I(1 + w_I) = 0,
\end{align}
where $w_I$ are the equation of state parameters associated with the matter species.

To investigate thermodynamical aspects of this cosmological model, we need to first associate the thermodynamic properties of the cosmological horizon, which is an apparent horizon for a general cosmological evolution. This necessitates the consideration of Hayward-Kodama formalism~\cite{Kodama:1979vn, Hayward:1997jp, Hayward:2008jq}: In an arbitrary spherically symmetric spacetime, $\d s^2=h_{ab}(x)\d x^a\d x^b+R^2(x)\d\Omega^2$, the apparent horizon is located where the gradient $\nabla_a R$ is null, i.e., $h^{ab}\partial_aR\partial_b R=0$. For a flat FLRW metric, the radius of apparent horizon becomes
\begin{align}
    R_{\textrm{AH}}(t)=\frac{1}{H}.
\end{align}
The temperature of a horizon is associated with the surface gravity $\kappa$, which is determined by solving the eigenvalue equation for the Killing vector field $\xi$ that generates the event horizon~\cite{Wald:1984rg}
\begin{align}
\xi^\mu\nabla_\mu\xi^\nu=\kappa\xi^\nu.
\end{align}
For dynamical spacetimes, the Kodama vector provides a convenient preferred timelike direction even though there is no timelike Killing vector \cite{Kodama:1979vn,Hayward:1997jp,Hayward:2008jq}. For the arbitrary spherically symmetric metric, the Kodama vector is 
\begin{align}
    K^a=\epsilon^{ab}\nabla_b R,
\end{align}
with $\epsilon^{ab}$ being the antisymmetric Levi-Civita tensor on the $t-r$ space. Just like the Killing vector for event horizons, Kodama vector is timelike in untrapped regions (outside the apparent horizon), spacelike in the trapped regions (inside the apparent horizon) and null on the marginally trapped apparent horizon \cite{Hayward:1997jp,Hayward:2008jq}. The notion of surface gravity for an apparent horizon is obtained by solving the eigenvalue equation for the Kodama vector
\begin{align}
    K^a\nabla_{[a} K_{b]}=\kappa_{\rm AH} K_b,
\end{align}
leading to
\begin{align}
    \kappa_{\textrm{AH}}=\frac{1}{2}\Box_hR\bigg|_{R=R_{\textrm{AH}}}.
\end{align}
For a flat FLRW universe, the surface gravity associated with the cosmological horizon takes the form \cite{Cai:2005ra}
\begin{align}
    \kappa_{\textrm{AH}}=-\frac{\dot{H}+2H^2}{2H}.
\end{align}
The Hayward-Kodama temperature \cite{Kodama:1979vn,Hayward:1997jp,Hayward:2008jq} associated with the cosmological horizon is given by
\begin{align}
\label{eq:temp}
T_{\textrm{AH}}=\frac{|\kappa_{\textrm{AH}}|}{2\pi}=\bigg|\frac{\dot{H}+2H^2}{4\pi H}\bigg|.
\end{align}
In the same spirit, the entropy of the apparent horizon is obtained from the area of apparent horizon ($A_{\textrm{AH}}$)
\begin{align}
    S_{\textrm{AH}}=2\pi A_{\textrm{AH}}=\frac{8\pi^2}{H^2}.
\end{align}
In a typical thermodynamical investigation of cosmological scenario, the matter content inside the horizon is assumed to be in thermodynamical equilibrium with the horizon \cite{Bousso:2004tv}. However, Mimoso and Pa\'{v}on showed that it is not the case for a radiation dominated universe, although thermalization is possible for non-relativistic matter and dark energy \cite{Mimoso:2016jwg}. Therefore, the conjecture of thermal equilibrium between horizon and matter content is expected to hold during the late-time evolution of the universe, particularly during the dark energy dominated era~\cite{Luciano:2022ffn, Duary:2019dfu,Duary:2023nnf,Chakrabarti:2024amb, Saha:2024mwn, Luongo:2012dv, Setare:2006vz, Gong:2006ma, Barboza:2015bha, Bhandari:2017cow,Bahamonde2018-qw,Chetry:2019tqd, Kong:2021dqd, Kong:2022xny, Abdusattar:2021wfv, Abdusattar:2023hlj, banihashemi2023minus,Sebastiani2023-sr,Cruz2023-sf,Panda2024-dg,Singh2024-wq,Cruz2024-iq, Chakrabarti:2025qsv,Samaddar_2025}.

It is, however, important to note the physical limitation of the equilibrium assumption. The horizon temperature today as per Eq.~\ref{eq:temp} is significantly lower than the temperature of the cosmic fluids (e.g., the CMB temperature today). This temperature gradient implies a heat flow between the fluid and the horizon. Physically, such an energy transfer would deform the homogeneous and isotropic FLRW background. Thus, treating the fluid and the horizon to have the same temperature is a working hypothesis to preserve the FLRW background, which is a potential caveat of this formalism (see~\cite{Jamil:2009eb} for a further discussion).

Accordingly, the entropy of the fluid inside the horizon $S_{\rm{in}}$ can be estimated from the first law of thermodynamics applied to this system
\begin{align}
    T_{\textrm{AH}}\d S_{\rm{in}}=\d U+P\d V,
\end{align}
where $V=4\pi R_{\textrm{AH}}^3/3=4\pi/3H^3$ is the volume of the fluid inside horizon and $U=\rho V$ is the internal energy of the fluid. The rate of change of entropy of fluid inside the horizon can be evaluated to (see, e.g.,~\cite{Duary:2023nnf})
\begin{align}
    \dot{S}_{\rm{in}}=\frac{32\pi^2\dot{H}(\dot{H}+H^2)}{H^3(\dot{H}+2H^2)},
\end{align}
while the rate of change of total entropy is
\begin{align}
    \dot{S}=\dot{S}_{\textrm{AH}}+\dot{S}_{\rm{in}}=\frac{16\pi^2\dot{H}^2}{H^3(\dot{H}+2H^2)}.
\end{align}
Here we see that as the horizon temperature vanishes for a universe with an effective equation of state $w=1/3$, the rate of change of entropy diverge. 
Although cosmological trajectories may transiently cross the $w=1/3$ boundary---a state where the Hayward-Kodama formalism yields unphysical results~\cite{Mimoso:2016jwg}---this limitation does not affect our primary conclusions. Here we focus on late-time stability and thermodynamic phase transitions. As we shall see, these phenomena and their corresponding dynamically stable future attractors are located in phase space domains well away from this crossing, thus the present analysis safely avoids this mathematical pathology.

The thermodynamical aspects of a system such as stability and second-order phase transition are typically characterized by the second derivatives of the thermodynamical potentials, i.e., specific heats at constant volume and constant pressure, $C_V, C_P$. For the flat FLRW universe, these quantities take the forms~\cite{Duary:2023nnf}
\begin{align}
\label{eq:cv-expr}
  C_V &= T\left(\frac{\partial S_{\rm in}}{\partial T}\right)_V =32\pi^2
        \sgn(2 H^2 + \dot{H})
        \frac{\dot{H}}{2H^2\dot{H}+H\ddot{H}-\dot{H}^2},\\
\label{eq:cp-expr}
   C_P &=  T\left(\frac{\partial S_{\rm in}}{\partial T}\right)_P = 32\pi^2
         \sgn(2 H^2 + \dot{H})
         \frac{H^2\dot{H}+\dot{H^2}}{H^2(2H^2\dot{H}+H\ddot{H}-\dot{H}^2)},
\end{align}
where $\sgn$ is the sign function coming from the absolute operator in the definition of temperature in Eq.~\ref{eq:temp}. Note that we assume $H>0$, thus focusing on the expanding branch of the universe. With the thermodynamic quantities associated with the FLRW universe at hand, we now briefly outline the requirements of phase transition and thermodynamic stability.
 
\emph{Phase transitions:} A thermodynamic system undergoes a first order phase transition if a first derivative of a thermodynamical potential (e.g., entropy, volume) is discontinuous, while a second-order phase transition is characterized by a discontinuity in the second derivative \cite{Blundell:2010}. Apart from the pathological discontinuity at radiation-dominated universe in $C_P, C_V$, the first order phase transitions are uncommon in the cosmological evolution. On the other hand, the second order phase transitions are natural to this setup~\cite{Duary:2023nnf,Chakrabarti:2024amb}, and in black hole thermodynamics; these are often termed as Davies-type phase transitions~\cite{Davies:1989ey}. Looking at the expressions of specific heats, it is apparent that the system undergoes second-order phase transitions when $2 H^2 + \dot{H}=0$ (finite discontinuity) and $2H^2\dot{H}+H\ddot{H}-\dot{H}^2=0$ (divergence). The first case is again pathological in nature and we will not consider it in this work. We shall focus on the second case where specific heats diverge, which has been investigated in different contexts earlier~\cite{Duary:2023nnf,Chakrabarti:2024amb}.
It is worth noting that if the specific heats were alternatively defined for the entire thermodynamic system using the total entropy $S_{\rm tot} = S_{\rm in} + S_{\rm AH}$, the values of the specific heats would naturally change. For instance, this would introduce an additional horizon contribution of the form $C_{V}^{\rm (AH)} = T_{\rm AH} \dot{S}_{\rm AH} / \dot{T}_{\rm AH}$. However, the divergence of these quantities is dictated by the extremum of the horizon temperature, where $\dot{T}_{\rm AH} \propto (2H^2\dot{H} + H\ddot{H} - \dot{H}^2)$ approaches zero. Because this term acts as a common denominator, the specific heats diverge at the exact same epoch regardless of the inclusion of the horizon entropy term.
 
\emph{Thermodynamic stability:} 
Following the treatment of~\cite{Callen:1985}, let us assume that the entire thermodynamic system consists of two identical subsystems, each described by an fundamental entropic relation $S \equiv S(U,V)$. Further suppose that the energy and the volume of the subsystems are perturbed as $(U, V) \to (U + \Delta U, V + \Delta V)$ and  $(U, V) \to (U - \Delta U, V - \Delta V)$, respectively. If the total system is thermodynamically stable, then such perturbations in the subsystems must not increase the total entropy. Therefore, thermodynamic stability requires that $2 S(U, V) \geq S(U + \Delta U, V + \Delta V) + S(U - \Delta U, V - \Delta V)$, which is mathematically equivalent of requiring that the entropy function $S(U, V)$ is concave in the $U-V$ space. 
These conditions in turn impose positivity of the specific heats at constant pressure and volume, and demand the difference between them be positive~\cite{Callen:1985,Barboza:2015bha}:
\begin{align}
\label{eq:stability_con}
C_P > 0,\quad C_V> 0,\quad C_P - C_V > 0. 
\end{align}
The positivity of specific heats alone ($C_P > 0, C_V> 0$) has significant cosmological implication. From the expressions of specific heats, we see their ratio is related to the deceleration parameter as
\begin{align}
    \frac{C_P}{C_V}=-q=1+\frac{\dot{H}}{H^2}.
\end{align}
which was first pointed out in~\cite{Saha:2024mwn}.\footnote{In \cite{Saha:2024mwn}, only $C_P>C_V$ is used as the sufficient criteria for thermodynamic stability while the positivity of individual specific heats is deemed immaterial.}
Therefore, for both specific heats to be positive --- which are necessary conditions for thermodynamic stability --- the deceleration parameter must be negative. Thus, \emph{the universe can be thermodynamically stable only during phases of accelerated expansion}, e.g., the late-time dark energy dominated epoch. Note that the converse is not guaranteed, i.e., accelerated expansion does not imply thermodynamic stability, as all the conditions in Eq.~\ref{eq:stability_con} may not be satisfied even if the specific heats have the same sign.


\section{Thermodynamical aspects on phase space}
\label{sec:thermo}

The key advantage of dynamical system analysis of cosmological models is that it can explore all viable evolutions of the universe, independent of the initial conditions.
Consequently, this approach allows one to identify the asymptotic fate of the universe in both past and future, and can characterize the transient phases. The dynamical system for the cosmological constant model and quintessence models of dark energy are extensively studied in the literature, see, e.g.,~\cite{Bahamonde:2018pr, Coley:2003mj, copeland2006dynamics} and references therein for further details. 
We should note that while standard dynamical system analysis evaluates stability at the background level, the inclusion of perturbations can sometimes alter the stability criteria of these critical points~\cite{Basilakos:2019dof}.
Here we briefly outline the aspects of the phase space analysis for the background evolution relevant to the present discussion. 

In general, a universe containing $n$ matter degrees of freedom is described by a phase space of $n-1$ dimensions, where each dimensionless phase space coordinate represents the fractional energy contribution of a specific matter component. The Hamiltonian constraint, or equivalently the Friedmann equation, reduces the dimensionality of the phase space by imposing an additional relation among these variables. The Raychaudhuri and the continuity equations of the matter sector leads to an autonomous system of linear differential equations that governs the evolution of the universe in the phase space. Given a set of initial energy budgets, the evolution of the universe is then represented as an orbit on this phase space.

Critical points in the phase space correspond to equilibrium states where the universe ceases to evolve. Stability analysis classifies these points as stable (attractor), unstable (repeller), or saddle points. Stable critical points attract nearby orbits, representing the universe's asymptotic future states, while unstable points act as past attractors from which orbits originate. Saddle points exhibit mixed behavior --- initially attracting and then repelling nearby trajectories --- and thus mark transient phases in cosmic evolution.

We map the thermodynamic quantities discussed in Sec.~\ref{sec:HK_temp} onto the phase space of cosmological models and establish a framework to: (i) identify potential thermodynamic phase transitions, and (ii) assess the thermodynamic stability of dark energy models.
Crucially, this approach shall be helpful in distinguishing thermodynamic aspects that are artifacts of initial conditions, from those intrinsic to the cosmological models themselves. We consider two scenarios: 
(i) the $\Lambda${}CDM concordance model, cosmological constant $\Lambda$ with subdominant contributions from non-relativistic matter and radiation, and
(ii) quintessence models of dark energy, where a dynamical scalar field drives the late-time acceleration.

\subsection{$\Lambda${}CDM model}
A detailed phase space analysis of the $\Lambda${}CDM model with non-relativistic matter (dust) and radiation is presented in~\cite{Bahamonde:2018pr,garcia2015introduction, fay2013ten}. Following these works, we define phase space coordinates for the $\Lambda${}CDM model as follows:
\begin{align}
  x &\equiv \Omega_M = \frac{\rho_M}{3 H^2},\\
  y &\equiv \Omega_R = \frac{\rho_R}{3 H^2},
\end{align}
where $\rho_M$ and $\rho_R$ are the energy densities of non-relativistic matter and radiation components.
The Friedmann equation constrains the density parameter of $\Lambda$ as
\begin{align}
  x + y + \Omega_\Lambda = 1.
\end{align}
The Friedmann equation and the positivity of the energy densities confine physical phase space in a triangle defined by $x,y \geq 0$ and $x + y \leq 1$.
The Raychaudhuri and the continuity equations of the matter species lead to the autonomous system of equations~\cite{Bahamonde:2018pr, garcia2015introduction}
\begin{align}
  x' &= x(3x + 4y - 3),\\
  y' &= y(3x + 4y - 4),
\end{align}
where prime denotes derivative with respect to the e-fold $\tau \equiv \ln a$. The autonomous system has three critical points, characterized by $(x'=0, y'=0)$:
\begin{enumerate}
\item $D\equiv(x=0,\; y=0)$: This is a stable critical point that behaves as a future attractor, describing a universe with only $\Lambda$. 
\item $R\equiv(x=0,\; y=1)$: This is an unstable critical point behaving as a past attractor, it describes a universe with only radiation. 
\item $M\equiv(x=1,\; y=0)$: This is a saddle point describing a universe with only non-relativistic matter. 
\end{enumerate}
Given the phase space description, we now recast the thermodynamic quantities in terms of the phase space variables.
The specific heat at constant volume ($C_V$) and constant pressure ($C_P$) are related to the phase space variables as:
\begin{align}
\label{eq:LCDM_cpcv}
  H^2 C_V &= \frac{64 \pi ^2 (3 x+4 y)}{9 x^2+24 x y-6 x+16 y^2-16 y},\\
    H^2 C_P &= -\frac{32 \pi ^2 (3 x+4 y-2) (3 x+4 y)}{9 x^2+24 x y-6 x+16 y^2-16 y}.
\end{align}
Note that we have dropped the $\sgn(2 H^2 + \dot{H}) = \sgn(1-q)$ factor from $C_V$ and $C_P$ (see Eqns.~\ref{eq:cv-expr} and~\ref{eq:cp-expr}), as within the boundary of the phase space ($0 \leq x, y \leq 1$ and $x + y \leq 1$), the factor is unity everywhere, $\sgn(1 - q) = \sgn \left(-\frac{3 x}{2}-2 y+2\right) = 1$. Recall that a second-order thermodynamic phase transition is characterized by the discontinuities in the specific heats. The above expressions indicate that the discontinuities occur along the curve in phase space defined by
\begin{align}
  \sigma(x,y) \equiv \frac{9 x^2}{4}+6 x y-\frac{3 x}{2}+4 y^2-4 y,
\end{align}
where the specific heats diverge. Therefore, an orbit in the phase space crossing this curve marks diverging specific heats and thus indicates a thermodynamic phase transition.

As discussed in Sec.~\ref{sec:thermo}, thermodynamic stability necessitates positivity of the specific heats, $C_V, C_P > 0$, and $C_P - C_V>0$. These requirements correspond to regions in the phase space, and their overlap describes domain where thermodynamical stability may occur. An orbit passing through such a domain at some time satisfies the necessary conditions for thermodynamic stability.

\begin{figure}
  \centering
  \includegraphics[width=.5\textwidth]{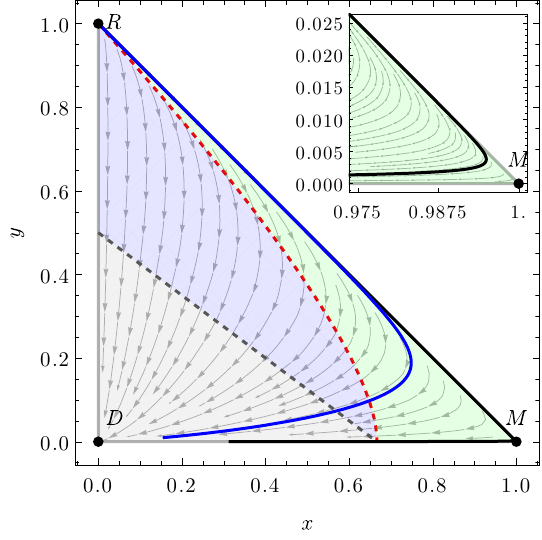}
  \caption{\emph{Phase space trajectories of the $\Lambda${}CDM universe.} The triangle bounded by $0 \leq x,y \leq 1$ and $ x + y \leq 1$ shows phase space with critical points $R$ (radiation domination), $M$ (matter domination), and $D$ (dark energy domination). Light gray arrows represent the slope field for all viable orbits. The dashed gray line marks the transition from deceleration to acceleration ($q=0$), with the gray shaded region indicating accelerated expansion. The red dashed line denotes the thermodynamic phase transition ($\sigma=0$). The green and blue regions highlight where $C_V>0$ and $C_P>0$, respectively. The black orbit traces the evolution of the $\Lambda${}CDM concordance model ($x(\tau=0) = \Omega_{M0} = 0.3153$, $y(\tau=0) = \Omega_{R0} = 10^{-4}$), and the blue orbit shows a case with reduced matter and enhanced radiation ($x(\tau=0) = \Omega_{M0}/2$, $y(\tau=0) = 10^2\Omega_{R0}$).}
  \label{fig:LCDMR}
\end{figure}
We plot the phase space trajectories of the $\Lambda${}CDM universe in Fig.~\ref{fig:LCDMR}, highlighting the regions of thermodynamic interest as follows:
The triangle marked by $0 \leq x,y \leq 1$ and $1 \leq x + y$ defines the phase space, where the critical points $R, M,$ and $D$ represent radiation, matter, and dark energy ($\Lambda$) dominated phases of the universe, respectively.
The light gray arrows represent the field of slopes for all viable orbits of the universe.
The dashed gray line marks a transition from deceleration to acceleration, $q = 0$, while the gray shaded region represents an accelerating universe.
The red dashed line marks the thermodynamic phase transition, $\sigma = 0$.
The green and blue shaded regions are where $C_V>0$, and $C_P>0$, respectively.

The black orbit denotes the evolution of the universe with a boundary condition consistent with observations~\cite{Planck:2020AA}: $x(\tau = 0) = \Omega_{M0} = 0.3153$, and $y(\tau = 0) = \Omega_{R0} = 10^{-4}$. Whereas, the blue orbit represents an example universe containing relatively lower matter contribution and higher radiation contribution at the current epoch, $x( \tau = 0) =\Omega_{M0} / 2  $, $y(\tau = 0) = 10^2 \Omega_{R0}$.

The unstable critical point $R$ acts as a past attractor, hence all orbits originate from a radiation-dominated era. These trajectories are subsequently drawn toward, and then repelled by, the saddle point $M$, representing a matter-dominated phase. How closely an orbit approaches the point $M$ depends on the relative matter content of the universe. Ultimately, all the orbits approach the future attractor $D$, corresponding to a dark energy–dominated de Sitter universe. Notably, the orbit corresponding to the concordance cosmological model (black curve) closely follows the trajectory $R \to M \to D$. In contrast, the blue orbit, representing a universe with lower matter content, is repelled by $M$ at an earlier time.

The phase space reveals a few interesting features. First, all orbits intersect the phase transition curve, this is evident from the phase space portrait (the only exception being the orbit $R \to D$ in the absence of matter). Although the $\sigma = 0$ curve originates at $R$, as do all the orbits, the curve never becomes tangent to any of the orbits. This follows from the fact that the directional derivative of $\sigma(x, y)$ along the slope field (say $\bm{v} \equiv \left\{x'(x,y), y'(x,y) \right\}$) is always non-zero along the curve $\d \sigma / \d \tau |_\sigma = \bm{v} \cdot \bm{\nabla} \sigma |_\sigma \neq 0$. 
Therefore, \emph{regardless of the relative energy budget, a spatially flat FLRW universe with $\Lambda$, matter, and radiation inevitably undergoes a phase transition indicated by diverging specific heats}. The only exception is the limiting case where the matter component is completely absent and the orbits lie entirely on the $y$-axis. We also see that the orbits with smaller matter contribution will undergo the phase transition at an earlier epoch.

Second, in this light, we may examine the interesting speculation raised in \cite{Duary:2023nnf}: whether such a phase transition is associated with the sign flip of the deceleration parameter $q$. To this end, first note that the curves $q = 0$ and $\sigma = 0$ do meet at the $x$-axis. Further, if we take $\Omega_R\rightarrow0$, which is effectively the model studied in~\cite{Duary:2023nnf}, it corresponds to orbit that connect $M \to D$ that lie on the $x$-axis. In this scenario, the events of the phase transition and the deceleration to acceleration transition \emph{indeed} coincide. However, even if a subdominant radiation component is included (the black orbit), the orbits although approach $M$ but never reach it. (see inset of Fig.~\ref{fig:LCDMR}). Therefore, these orbits never lie on the $x$-axis hence they encounter the phase transition and the deceleration-acceleration transition at different times. This difference is more obvious for trajectories with relatively less matter components (the blue curve). Therefore, although the phase transition and signature flip of $q$ are \emph{almost simultaneous} for the concordance model of cosmology, these two phenomenon are not fundamentally connected. Notably, the thermodynamic phase transition occurs in the decelerating phase when non-zero radiation component is present.

Third, we observe that the regions where the specific heats are positive ($C_V > 0$: green region and $C_P>0$: blue region) have no intersection. In fact, these regions are separated by the phase transition curve, where these quantities diverge. This is an artifact of $\Lambda${}CDM evolution and not a generic feature as we will see in the next subsection. Consequently, a necessary condition for thermodynamic stability is never met in the entire phase space. Thus, this exercise, with the radiation component included, generalizes the observations by~\cite{Duary:2023nnf} for the $\Lambda$ plus dust universe. Interestingly, even though the future attractor $D$ is a dynamically stable critical point, it lies in a region where both the specific heats are negative. Therefore, the notion of thermodynamic stability --- at least as build upon the Hayward-Kodama formalism--- is at odds with the notion of dynamical stability associated with the evolution of the $\Lambda${}CDM universe.

\begin{figure}
  \centering
  \includegraphics[width=.5\textwidth]{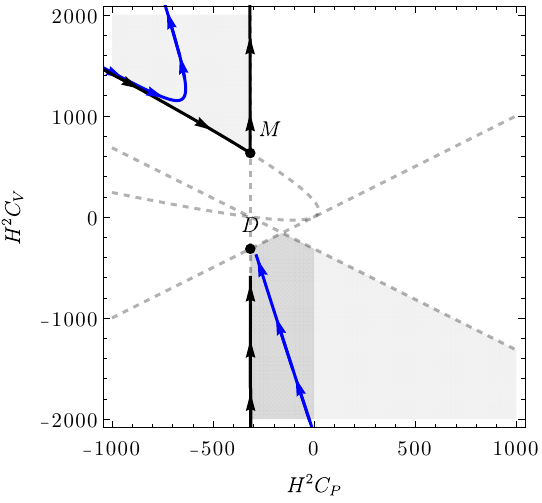}
  \caption{
  Physical phase space of the $\Lambda$CDM model mapped onto the specific heat ($H^2 C_P - H^2 C_V$) plane. The physical bounds ($0 \le x, y \le 1$ and $x+y \le 1$) map to two disjoint shaded patches bounded by the dashed gray curves. The dark shaded region marks accelerated expansion. The physical regions never intersect the first quadrant ($C_P > 0, C_V > 0$), showing that the necessary conditions for thermodynamic stability are never simultaneously satisfied. The solid blue and black curves represent the same cosmic orbits shown in Fig.~\ref{fig:LCDMR}. The trajectories emerge from the unstable radiation node $R$ (located at infinity in the second quadrant), approach the matter-dominated saddle point $M$ and encounter a thermodynamic phase transition where the specific heats diverge. The trajectories consequently diverge to infinity, re-entering from the third quadrant, before finally converging at the stable, dark energy-dominated future attractor $D$.}
  \label{fig:CP-CV_plane}
\end{figure}

It is interesting to observe the evolution of the $\Lambda$CDM universe directly in the $C_P-C_V$ plane.
To this end, we invert the expressions for the specific heats in Eq.~\eqref{eq:LCDM_cpcv} to write the phase space variables in terms of them as 
\begin{align}
   x(\tilde{C}_P, \tilde{C}_V) 
   &= 
   -\frac{2 \left(\tilde{C}_P-\tilde{C}_V\right) \left(\tilde{C}_P+\tilde{C}_V+32 \pi ^2\right)}{3 \tilde{C}_V^2},\\
    y(\tilde{C}_P, \tilde{C}_V) 
   &= 
   \frac{\left(\tilde{C}_P+32 \pi ^2\right) \left(\tilde{C}_P-\tilde{C}_V\right)}{2 \tilde{C}_V^2},
\end{align}
where $\tilde{C}_P \equiv H^2 {C}_P$ and $\tilde{C}_V \equiv H^2 {C}_V$. 
Figure~\ref{fig:CP-CV_plane} depicts the physical phase space of the $\Lambda$CDM model in the $C_P-C_V$ space, marked by the shaded regions. 
The boundaries of the phase space ($0 \leq x, y \leq 1$ and $x + y \leq 1$) translate into the inequalities 
$-2 \left(\tilde{C}_P-\tilde{C}_V\right) \left(\tilde{C}_P+\tilde{C}_V+32\pi ^2\right) \geq 0$, 
$\left(\tilde{C}_P+32 \pi ^2\right)\left(\tilde{C}_P-\tilde{C}_V\right) \geq 0$, 
and $\left(\tilde{C}_V-\tilde{C}_P\right) \left(\tilde{C}_P+4 \left(\tilde{C}_V+8\pi ^2\right)\right)-6 \tilde{C}_V^2 \leq 0$
in the specific heat plane (denoted by the dashed gray curves). 
As is evident, the physical phase space is mapped to two disjoint patches that extend to infinity along both axes.
Crucially, these patches never intersect in the first quadrant where both $C_P > 0$ and $C_V > 0$, thus showing that the necessary conditions for thermodynamic stability are never simultaneously satisfied.
We further plot the same two orbits of the universe as in Fig.~\ref{fig:LCDMR} (depicted by the solid blue and black curves). 
The trajectories emerge from the unstable radiation node $R$ (located at infinity in the second quadrant) and evolve toward the matter-dominated saddle point $M$. 
The evolutions then encounter the thermodynamic phase transition, where the specific heats diverge. Consequently, the trajectories shoot off to infinity, re-entering the plane from the opposite asymptotic limit in the third quadrant before ultimately converging at the stable, dark-energy-dominated future attractor $D$.

\subsection{Quintessence models}

We now extend the discussion to quintessence models of dark energy: a scalar field ($\phi$) minimally coupled to gravity with a canonical kinetic term (see, for example, \cite{tsujikawa2013quintessence} and references therein). Phase space analysis of various quintessence models are extensively explored in the literature~\cite{faraoni2013scalar, Bahamonde:2018pr, amendola2010dark,copeland2006dynamics,Copeland:1998PhRvD,Urena-Lopez:2012JCAP,Halliwell:1987PhLB, Steinhardt:1999PhRvD,Macorra:2000PhRvD, Ng:2001PhRvD, Scherrer:2008PhRvD, Gong:2014PhLB, roy2014tracking,Tamanini:2014PhRvD, Singh:2015rqa, Qi:2016IJTP}. Typically, the kinetic energy and the potential energy contributions of the scalar fields, $\dot{\phi}^2 / 2$ , $V(\phi)$ are treated as distinct dynamical variables. Furthermore, for a general potential term, the system of equations must also include the evolution equation for the potential itself, leading to a higher-dimensional phase space~\cite{Steinhardt:1999PhRvD, Macorra:2000PhRvD, Ng:2001PhRvD}. In this work, we restrict our discussion to the simple case of exponential potentials~\cite{Bahamonde:2018pr, amendola2010dark,Copeland:1998PhRvD, Urena-Lopez:2012JCAP,Burd:1988NuPhB}
\begin{align}
  V(\phi) = V_0 \exp(- \lambda  \phi ).\label{expPot}
\end{align}
The model parameter $\lambda$ plays a crucial role in determining the critical points and the fate of the universe, whereas the constant $V_0$ does not influence the phase space dynamics. For this choice of potential, the system of equations becomes autonomous and the phase space closes. Additionally, we include a non-relativistic matter component ($\rho_M$) and neglect the contribution from radiation in order to restrict the phase space to two dimensions.

The phase space variables $x,y$ are taken to be the square roots of the kinetic and potential energy contributions of the scalar field, respectively~(see, e.g.,~\cite{Bahamonde:2018pr, amendola2010dark}):
\begin{align}
  x &\equiv \frac{\dot{\phi}}{\sqrt{6}H},\\
  y &\equiv \frac{\sqrt{V}}{\sqrt{3} H}.
\end{align}
The Friedmann equation constrains the density parameter of matter $\Omega_M = \rho_M/(3 H^2)$ as
\begin{align}
  x^2 + y^2 + \Omega_M = 1.
\end{align}
The Friedmann equation and the positivity of the energy densities confine phase space in a half-circle defined by $ y \geq 0$ and $x^2 + y^2 \leq 1$.
The Raychaudhuri equation, the continuity equation for matter, and the Klein-Gordon equation for the scalar field lead to the autonomous system of equations~\cite{Bahamonde:2018pr, amendola2010dark} 
\begin{align}
  x' &= -\frac{3}{2} \left(-x^3+x \left(y^2+1\right)-\sqrt{\frac{2}{3}} \lambda 
       y^2\right),\\
  y' &= -\frac{3}{2} y \left(-x^2+\sqrt{\frac{2}{3}} \lambda  x+y^2-1\right).
\end{align}
The autonomous system has four critical points:
\begin{enumerate}
\item $K_{\pm} \equiv (x = \pm 1,\; y = 0)$:  
  For $-\sqrt{6} \leq \lambda \leq \sqrt{6}$, these points are unstable nodes and act as past attractors for all orbits. They correspond to scalar field–dominated universes, where the energy density of the field is purely kinetic. Physically, this mimics a stiff fluid with an equation of state $1$.

\item $M \equiv (x = 0,\; y = 0)$:  
This is a saddle point representing a matter-dominated universe. It attracts orbits emerging from the kinetic-dominated points, subsequently repels  them to stable nodes.

\item $S \equiv \left(x = \sqrt{\frac{3}{2}} \frac{1}{\lambda},\; y = \sqrt{\frac{3}{2}} \frac{1}{\lambda} \right)$:  
  This point exists for $\lambda^2 \geq 3$ and is a stable critical point, acting as a future attractor. It corresponds to a \emph{scaling solution} where the scalar field's equation of state mimics that of the matter component. This behavior is phenomenologically interesting as it can address the coincidence problem. However, the universe is decelerating at this stable node since the effective equation of state remains $0$ here.
\item $\Phi \equiv \left(x = \frac{\lambda}{\sqrt{6}},\; y = \sqrt{1 - \frac{\lambda^2}{6}} \right)$:  
This point exists for $\lambda^2 < 6$. It behaves as a stable node (future attractor) when $\lambda^2 < 3$, and as a saddle point when $3 \leq \lambda^2 < 6$. Physically, it represents a scalar field–dominated universe, where both kinetic and potential contributions are present and the universe accelerates if $\lambda^2 < 2$.
\end{enumerate}
In this case, the specific heats are related to the phase space variables as:
\begin{align}
  H^2 C_V &=
            \sgn(1 - q)
\frac{48 \pi ^2 \left(x^2-y^2+1\right) }{\sigma(x,y)},\\
  H^2 C_P &= - \sgn(1 - q)
            \frac{24 \pi ^2 \left(3 x^2-3 y^2+1\right) \left(x^2-y^2+1\right)}{\sigma(x,y)},
\end{align}
where the function $\sigma$ is now defined as
\begin{align}
  \sigma(x,y) \equiv
  \frac{1}{4} \left(9 x^4-6 y^2 \left(3 x^2-2 \sqrt{6} \lambda  x+2\right)-24
   x^2+9 y^4+3\right).
\end{align}
Again, the phase transition indicated by diverging specific heats occur along the curve $\sigma = 0$ in phase space. It is also convenient to note the form of the deceleration parameter in the phase space,
\begin{align}
  q(x,y) = \frac{1}{2} \left(3 x^2-3 y^2+1\right).
\end{align}
In this model, the factor $\sgn(1-q)$ in the specific heats becomes important --- as the universe approaches the regime of kinetic energy domination, the deceleration parameter may decrease below $1$ thus changing the sign of this factor. 

Before turning to the specifics of the models, we may infer the possibility of thermodynamic stability from the analytical expressions above. First recall that since $C_P/C_V = -q$ \cite{Saha:2024mwn}, both $C_P$ and $C_V$ can become positive only where the universe is accelerating.
Let us assume that the universe is at such an accelerating region in the phase space, thus $
\sgn(1 - q) = 1$.
Furthermore, since the equation of state of the quintessence field does not cross the phantom divide, i.e. $w_\phi > -1$, the deceleration parameter is restricted as $ q > -1$, implying $x^2 - y^2 + 1 > 0$. Given these, the requirement $C_V > 0$ thus implies $\sigma(x, y) > 0$.
On the other hand, within these regions, the difference between the specific heats can be put in the form
\begin{align}
  C_P - C_V = - \frac{72 \pi ^2 \left(x^2-y^2+1\right)^2 }{\sigma(x,y)}.
\end{align}
Thus the requirement  $C_P - C_V >0$ implies $\sigma(x, y) < 0$. Therefore, we see that the three necessary conditions for thermodynamic stability --- $C_V >0$, $C_P>0$, and $C_P-C_V >0$ --- \emph{are never to be  satisfied simultaneously anywhere in the phase space, rendering these quintessence models thermodynamically unstable regardless of their initial configurations.}

To demonstrate this explicitly, let us consider two examples of the quintessence models at hand:
\subsubsection{$\lambda = 1$: accelerating future}
This choice of parameter leads to three critical points in the phase space: $K_{\pm}, M, \Phi$, while the point corresponding to scaling solution $S$ is absent.
\begin{figure}
  \centering
  \begin{subfigure}{0.49\textwidth}
    \centering
    \includegraphics[width=\textwidth]{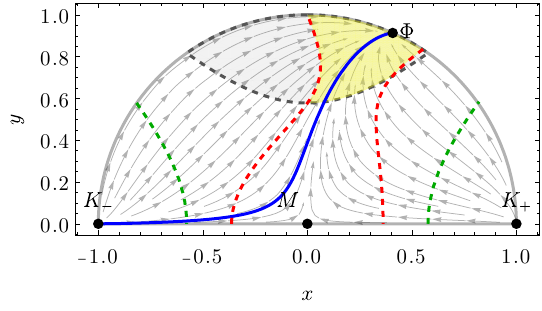}
  \caption{Phase space}    \label{fig:Q1a}
  \end{subfigure}
  \begin{subfigure}{0.49\textwidth}
    \centering
    \includegraphics[width=\textwidth]{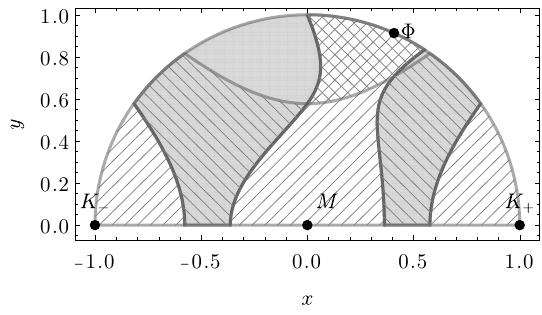}
    \caption{Thermodynamic regions}
    \label{fig:Q1b}
  \end{subfigure}
  \caption{\emph{Phase space of the quintessence model with exponential potential, $\lambda =1$.} The half-circle bounded by $x^2 + y^2 \leq 1$ and $y>0$ shows the phase space and thermodynamic regions in panels (a) and (b), respectively. Both panels show the critical points $K_{\pm}$ (Kinetic domination), $M$ (matter domination), and $\Phi$ (scalar field–domination). (a) \emph{Phase space:} The light gray arrows show the slope field of the orbits. The dashed gray line marks the transition from deceleration to acceleration ($q=0$), with the gray shaded region indicating accelerated expansion. The green dashed line marks $q = 1$. The red dashed line denotes the thermodynamic phase transition ($\sigma=0$). Yellow region indicates the overlap of $C_V>0$ and $C_P>0$. The blue orbit shows cosmic evolution with boundary condition $x(\tau = 0)=0.2124$, $y(\tau = 0)=0.7888$ which yields $\Omega_\phi(\tau = 0) \approx 0.667$ at present. (b) \emph{Thermodynamic regions:} The panel shows regions for three thermodynamic stability conditions, {with forward and backward hatching representing $C_V>0$ and $C_P>0$, respectively.} The gray shaded region shows $C_P - C_V > 0$.}
  \label{fig:Q1}
\end{figure}
In Fig.~\ref{fig:Q1a}, the gray arrows represent the slope field of the autonomous system. The gray shaded region bounded by dashed gray lines corresponds to accelerated expansion of the universe, where $q(x, y) < 0$, while the yellow shaded region marks the domain where both $C_P > 0$ and $C_V > 0$ are satisfied simultaneously. The red, gray, and green dashed lines indicate, respectively, the phase transition curve $\sigma(x, y) = 0$, the boundary between deceleration and acceleration, $q(x, y) = 0$, and radiation dominated epoch $q(x,y)=1$.
The solid blue curve shows an orbit of the universe with the boundary condition $x(\tau = 0) = 0.2124$ and $y(\tau = 0) = 0.7888$, yielding a present-day scalar field density parameter of $\Omega_\phi(\tau = 0) \approx 0.667$.

The unstable critical points $K_{\pm}$ act as past attractors, all orbits originate from an era where the kinetic energy of the quintessence field dominates. These trajectories are subsequently drawn toward, and then repelled by, the matter-dominated saddle point $M$, Eventually, all the orbits approach the future attractor $\Phi$ where the potential energy of the scalar field dominates and the universe is accelerating.

Interestingly, there are now two distinct phase transition curves; they separates the future attractor $\Phi$ from the two past attractor $K_-$ and $K_+$. As a result, all the orbits originating from either $K_-$ or $K_+$ --- depending on the initial velocity of the scalar field --- must cross one of these phase transition curves to reach the future attractor $\Phi$. Thus, a thermodynamic phase transition is  inevitable for this quintessence model, regardless of the initial conditions. Moreover, it is evident from the phase space portrait that the thermodynamic phase transition does not occur during or before the pathological case of a radiation-dominated universe,  as the orbits encounter the $\sigma = 0$ curve after the $q=1$ curve.

Furthermore, the curve $q(x, y) = 0$ does not lie on either of the phase transition curves, although it intersects both. Consequently, the phase transition and the deceleration-acceleration transition generally occur at different times along a given orbit, indicating that these events are unrelated.  However, for some finely tuned initial conditions,  the orbits may intersect the $q(x,y) = 0$ and $\sigma(x, y) = 0$ curves simultaneously.

Figure~\ref{fig:Q1b} highlights the regions satisfying the three thermodynamic stability conditions: {$C_V > 0$ is indicated by forward hatching, $C_P > 0$ by backward hatching}, and $C_P - C_V > 0$ by the gray shaded region. In contrast to the $\Lambda${}CDM model, there exits a region where both the specific heats are positive and the dynamically stable future attractor $\Phi$ lies within this region. However, as noted before and shown in Fig.~\ref{fig:Q1b}, there exists no intersection among all the three regions --- $C_P>0$, $C_V>0$, and $C_P - C_V>0$; hence the  necessary conditions for thermodynamic stability is never satisfied in the phase space, similar to that of the  $\Lambda${}CDM model.

\subsubsection{$\lambda = 2$: decelerating future and scaling solution}
As another example, we are considering the parameter choice that leads to four critical points: $K_{\pm}, M, \Phi$, and $S$. In this case, the future attractor is the scaling solution $S$, where the universe is driven by both the quintessence field and matter. However, the equation of state of the quintessence field mimics that of matter, resulting in a decelerating universe. Here  the point $\Phi$ becomes a saddle, while the natures of $K_\pm$ and $M$ remains the same as before.
\begin{figure}
  \centering
  \begin{subfigure}{0.49\textwidth}
    \centering
    \includegraphics[width=\textwidth]{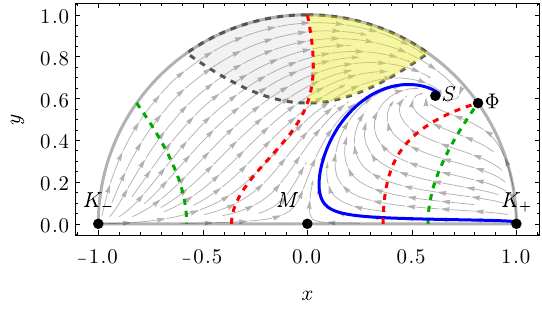}
    \caption{Phase space}
    \label{fig:Q2a}
  \end{subfigure}
  \begin{subfigure}{0.49\textwidth}
    \centering
    \includegraphics[width=\textwidth]{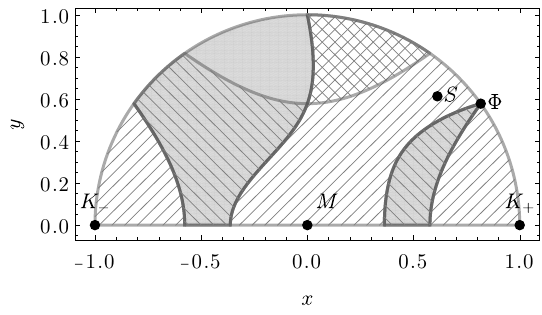}
    \caption{Thermodynamic regions}
    \label{fig:Q2b}
  \end{subfigure}
  \caption{\emph{Phase space of the quintessence model with exponential potential, $\lambda =2$.} Format and convention follow Fig.~\ref{fig:Q1}, apart from the additional critical point $S$ (scaling solution) appearing in the case. The blue orbit in panel (a) shows cosmic evolution with boundary condition $x(\tau = 0)=y(\tau = 0)=0.6123$ which yields $\Omega_\phi(\tau = 0) \approx 0.75$ at present.}
  \label{fig:Q2}
\end{figure}
We depict the phase space for this model in Fig.~\ref{fig:Q2}, where the interpretations of the curves and the shaded regions remain the same as in Fig.~\ref{fig:Q1}. The orbit (solid blue curve) is evolved from the boundary condition $x(\tau = 0) = 0.6123$ and $y(\tau = 0) = 0.6123$, resulting in a current scalar field density parameter of $\Omega_\phi(\tau = 0) \approx 0.75$.

The key difference between this model and the previous one is that, here the universe finally settles to a decelerating phase. Depending on the initial conditions, there still may exist transient phases of acceleration. However, due to the choice of the initial condition, the blue orbit avoids any phase of acceleration. As before, we observe two distinct phase transition curves that separate the future attractor $S$ from the two past attractors $K_-$ and $K_+$. However, one of these curves terminate at the future saddle $\Phi$ in this case. Thus, all trajectories starting from either $K_-$ or $K_+$ must intersect these curves to reach the future attractor $S$, indicating an inevitable phase transition.
We again observe that the phase transition and the deceleration-acceleration transition events are fundamentally unrelated. For example, although the blue orbit undergoes a phase transition, the universe remains decelerating throughout along this orbit.

Figure~\ref{fig:Q2b} highlights the regions satisfying the three thermodynamic stability conditions, where the meanings of the different region legends remain the same as in Fig.~\ref{fig:Q1b}. As in the previous case, there is no region in the phase space where all the three stability conditions are simultaneously satisfied, however, both specific heats are positive during the accelerated expansion. Here, the stable future attractor $S$ lies in a region of decelerated expansion where only {$C_V>0$}. 
As the dynamically stable future attractors can lie in regions where either or both specific heats are negative, these observationally favored cosmological models do not indicate a generic thermodynamic feature.
Here we should note that while it would be interesting to see the evolution of the universe directly in the $C_P-C_V$ plane, as we saw in the case of the $\Lambda$CDM model in Fig.~\ref{fig:CP-CV_plane}, this is not feasible for these cases. This is because the relations between the specific heats and the phase space variables become highly non-linear for the scalar field models, which renders the inversions for $x(C_P,C_V)$ and $y(C_P,C_V)$ intractable.
\subsection{Phantom models}
\label{sec:phantom}
As we see from our previous discussions, both the concordance model and quintessence models fail to achieve thermodynamic stability. A common feature of these scenarios is that the effective equation of state parameter is confined to the regime $w \geq -1$. Given this limitation, we ask whether thermodynamic stability might be attainable in models where the equation of state is allowed to cross the phantom divide $w < -1$, as this extends the domain of the phase space. To investigate this possibility, we consider a phantom dark energy model, where unlike quintessence, the sign of the kinetic term in the phantom Lagrangian is reversed~\cite{Caldwell:1999ew,Caldwell:2003vq}:
\begin{align}
    \mathcal{L}_\phi=\frac{1}{2}\partial_\mu\phi\partial^\mu\phi-V(\phi).
\end{align}
The equation of state for this scalar field can attain $w_\phi<-1$ leading to the phantom regime in the evolution of the universe, thus known as a phantom field. Phantom regime is phenomenologically intriguing as it is favored by the astronomical observation, albeit not with statistical significance~\cite{Valentino:2020JCAP,DESI:2025JCAP,Chakraborty:2025JCAP}. The non-canonical fields carry many theoretical issues, notably perturbative instability classically and spectrum instability at quantum level~\cite{Carroll:2003st,Cline:2003gs}. Therefore, these models are treated as emergent phenomenon which should not be trusted at fundamental level, akin to the thermodynamic aspects of cosmological evolution.

\begin{figure}
  \centering
  \begin{subfigure}{0.49\textwidth}
    \centering
    \includegraphics[width=0.99\textwidth]{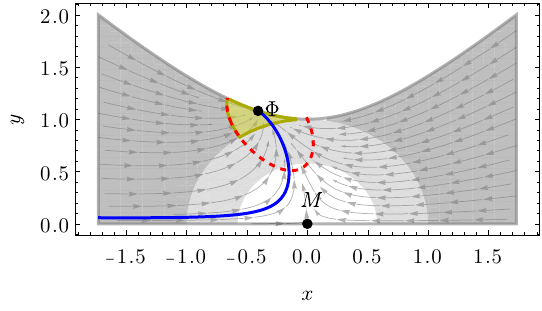}
  \caption{Phase space}    \label{fig:P1a}
  \end{subfigure}
  \begin{subfigure}{0.49\textwidth}
    \centering
    \includegraphics[width=0.99\textwidth]{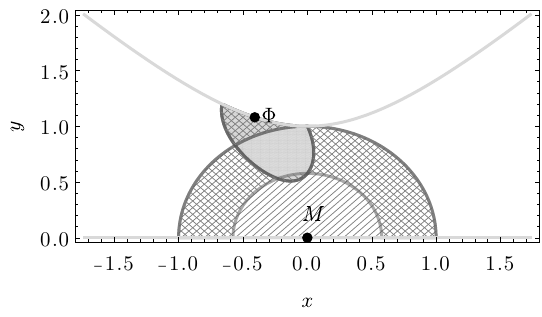}
    \caption{Thermodynamic regions}
    \label{fig:P1b}
  \end{subfigure}
  \caption{\emph{Phase space of the phantom model with exponential potential, $\lambda =1$.} The region bounded by hyperbola $-x^2 + y^2 \leq 1$ and $y>0$ shows the phase space and thermodynamic regions in panels (a) and (b), respectively. Both panels show the critical points, $M$ (matter domination), and $\Phi$ (scalar field–domination) while the critical points for Kinetic domination are at the infinities in the $(x,y)$ plane. (a) \emph{Phase space:} The light gray arrows show the slope field of the orbits. The gray shaded region indicate the accelerated expansion: light gray corresponds to the standard accelerated expansion $-1<w_{\rm eff}<-1/3$, while the dark gray region corresponds to the phantom regime $w_{\rm eff}<-1$. The red dashed curve denotes the thermodynamic phase transition ($\sigma=0$). The yellow region indicates the overlap of $C_V>0$, $C_P>0$, and $C_P-C_V>0$, i.e., region of thermodynamic stability. The blue orbit shows cosmic evolution with a boundary condition $x(\tau = 0)=-0.212$, $y(\tau = 0)=0.788$, which yields $\Omega_\phi(\tau = 0) \approx 0.576$ at present. (b) \emph{Thermodynamic regions:} The panel shows regions for three thermodynamic stability conditions, with forward and backward hatching representing $C_V>0$ and $C_P>0$, respectively. The gray shaded region shows $C_P - C_V > 0$.}
  \label{fig:P}
\end{figure}

Again, restricting our analysis to the simple case of exponential potentials Eq.~\ref{expPot}, the phase space variables of this model are taken to be the square roots of kinetic and potential energy contributions of the scalar field~\cite{Bahamonde:2018pr}
\begin{align}
    x=\frac{\dot{\phi}}{\sqrt{6}H},\quad y=\frac{\sqrt{V}}{\sqrt{3}H}.
\end{align}
With the pressureless dust, the Friedmann equation constraint the phase space according to
\begin{align}
    -x^2+y^2+\Omega_M=1.
\end{align}
Since $\Omega_M$ is positive, the physical phase space is confined to the region bounded by the hyperbola $-x^2+y^2<1$ and $y>0$ as shown in Fig.~\ref{fig:P}. The Raychaudhuri equation and Klein-Gordon equation for the scalar field lead to the autonomous system of equations for this system \cite{Bahamonde:2018pr}
\begin{align}
    x'=&-\frac{3}{2}\left(x^3+x(y^2+1)+\sqrt{\frac{2}{3}}\lambda y^2\right),\\
    y'=&-\frac{3}{2}y\left(x^2 +  (y^2-1) + \sqrt{\frac{2}{3}}\lambda x\right).
\end{align}
The critical points in the system are
\begin{enumerate}
\item $M \equiv (x = 0 ,\; y = 0)$:  
 This is a saddle point representing a matter-dominated universe. It attracts orbits emerging from the past attractors, subsequently repelling  them to stable nodes.

\item $\Phi \equiv (x = {-} \lambda/\sqrt{6},\; y = \sqrt{1 + \lambda^2 / 6})$:  
This is a stable point representing a phantom field-dominated universe. The universe is accelerating at this point.
\end{enumerate}
The specific heats as functions of phase space variable become
\begin{align}
  H^2 C_V &=
             - \sgn(1 - q)
\frac{48 \pi ^2 \left(x^2 + y^2 - 1\right) }{\sigma(x,y)},\\
  H^2 C_P &= - \sgn(1 - q)
            \frac{24 \pi ^2 \left(3 x^2 + 3 y^2 -1\right) \left(x^2 + y^2 - 1\right)}{\sigma(x,y)},
\end{align}
where the function $\sigma$ is for the phantom model is defined as
\begin{align}
  \sigma(x,y) \equiv
\frac{1}{4} \left(9 x^4+6 y^2 \left(3 x^2+2 \sqrt{6} \lambda 
   x-2\right)+24 x^2+9 y^4+3\right)
\end{align}
Note that for phantom models, the factor $\sgn(1-q)$ becomes $\sgn(1 + 3 x^2 + 3 y^2)$, thus remains $\sgn(1-q) = 1$ throughout the phase space. 
To infer the viability of phase space stability in phantom models, let us assume that the universe is at a phantom region in the phase space, $q < -1$ implying $x^2 + y^2 - 1 > 0$. Now if further $\sigma (x, y) < 0$ in the region as well, we may infer from the above expressions that $C_P, C_V > 0$ would be satisfied. 
On the other hand, within this region, the difference between the specific heats can be put in the form
\begin{align}
  C_P - C_V = - \frac{72 \pi ^2 \left(x^2 + y^2 - 1\right)^2 }{\sigma(x,y)}.
\end{align}
Thus the requirement  $C_P - C_V >0$ is satisfied as $\sigma(x, y) < 0$, which is compatible with the other necessary conditions in this case. Therefore, given there exists a region in the phase space where $\sigma(x, y) < 0$, the three necessary conditions for thermodynamic stability --- $C_V >0$, $C_P>0$, and $C_P-C_V >0$ --- may be satisfied simultaneously. This indicates that \emph{phantom models may achieve thermodynamical stability}, unlike the quintessence models discussed above.

We depict the phase space for the phantom model ($\lambda=1$) in Fig.~\ref{fig:P}. Crucially, the yellow shaded region identifies the domain where the universe is thermodynamically stable, satisfying $C_V > 0$, $C_P > 0$, and $C_P - C_V > 0$ simultaneously. Moreover, the future attractor $\Phi$ lies within this region of thermodynamic stability.
The solid blue curve traces a representative orbit of the universe with the boundary condition  
$x(\tau = 0)=-0.212$, $y(\tau = 0)=0.788$, corresponding to a present-day scalar field density parameter of $\Omega_{\phi}(\tau=0) \approx 0.576$. The red dashed line marks the thermodynamic phase transition curve defined by $\sigma(x,y)=0$. As the line of phase transition encloses the future attractor, all trajectories in the phase space undergo a phase transition for the phantom models.

This observation is particularly intriguing; phantom scalar fields, despite suffering from dynamical instabilities under small perturbations~\cite{Carroll:2003st,Cline:2003gs}, are capable of attaining thermodynamic stability at the future attractor. 
A crucial ingredient for achieving this stability is the evolution of the deceleration parameter into the regime $q < -1$, a feature characteristic of phantom models. 
This stands in contrast to the quintessence and $\Lambda$CDM cases, where the future attractors are thermodynamically unstable.

\section{Summary and discussion}
\label{sec:summary}

We investigate thermodynamic aspects of the cosmological evolution of the spatially flat FLRW universe using the Hayward-Kodama formalism. We focus on three scenarios: the
$\Lambda${}CDM concordance model of cosmology, the quintessence models of dark energy with exponential potentials, and phantom model of non-canonical scalar field with exponential potential.
We recast the relevant thermodynamic quantities directly into the phase space coordinates. This mapping provides valuable insights into their global behavior, particularly at the critical points. Consequently, we can evaluate thermodynamic stability and phase transitions in a way that is independent of the initial conditions of the evolution.

In all the models under consideration, the phase transition curve --- indicated by diverging specific heats --- appears to create a separation between the past and future attractors in the phase space. Consequently, all orbits corresponding to physically viable solutions must intersect these curves at some stage of the evolution. This exercise indicates that \emph{a thermodynamic phase transition is thus inevitable during the evolution of the universe} and not an artifact of the choice of the initial condition.

Addressing a speculation raised in~\cite{Duary:2023nnf}, we show that the deceleration to acceleration transition is generally distinct from the thermodynamic phase transition. Although the two transitions may occur `nearly simultaneously' for the concordance model, these phenomena are not fundamentally connected. In fact, the phase transition occurs during the decelerating phase for $\Lambda${}CDM model with finite radiation component. This distinction is more obvious in the case of quintessence models, where the universe might never accelerate but encounters a phase transition for some initial conditions, while it can also accommodate phase transition during accelerated phase for other initial conditions.

From the general relation between specific heats and deceleration parameter, $C_P/C_V=-q$, it is clear that only an accelerating universe can sustain positive specific heats, i.e., thermodynamic stability. However, the necessary criteria of thermodynamic stability, $C_P>0$, $C_V>0$, and $C_P - C_V>0$, are never simultaneously satisfied anywhere in the phase space for the concordance model and quintessence models, therefore the evolution of the universe remains thermodynamically unstable. In particular, at least one of these stability conditions is violated for the dynamically stable future attractors. 
We emphasize, however, that this lack of concavity in the entropy function does not imply the ruling out of the $\Lambda${}CDM or quintessence background models. The present results could also mean that the standard canonical ensemble based stability criteria may be limited, or inadequate, when applied to cosmological horizons.

Finally, we extended our analysis to phantom dark energy models, where the equation of state parameter is permitted to cross into the regime $w< -1$. Similar to the previous models, we found that a thermodynamic phase transition --- marked by diverging specific heats --- remains a generic feature of the evolution, as the transition curve encloses the future attractor in the phase space. However, the thermodynamic stability of the phantom models differs significantly. We show that access to the phantom regime allows all necessary conditions for thermodynamic stability to be satisfied simultaneously, $C_P>0$, $C_V>0$, and $C_P - C_V>0$. Notably, the future attractor of the phantom models lies within this region of thermodynamic stability. This leads to an intriguing observation: while phantom scalar fields typically suffer from dynamical instabilities under small perturbations \cite{Carroll:2003st,Cline:2003gs}, they uniquely attain thermodynamic stability at their asymptotic future. 

In conclusion, our analysis demonstrates that the dynamical systems approach is particularly effective for uncovering thermodynamic features of cosmological evolution. It is interesting to check whether our observation of thermodynamical stability in phantom regime is a model independent artifact. Furthermore, it is worthwhile to investigate if crossing the phantom divide is necessary to attain a thermodynamic stability. The framework that we have introduced in this work is perfect to address these questions.

On a cautionary note, we admit the notion of thermodynamic stability in gravitational scenarios is a delicate issue, as one has to be cognizant of the long-range nature of gravity \cite{Padmanabhan:1989gm}, which is reflected in the negative specific heats for these systems~\cite{HERTEL1971520}. 
In the standard canonical ensemble framework, these thermodynamic systems are unstable~\cite{Callen:1985}. However, it has been argued that the stability of gravitational systems should be treated in the microcanonical framework where negative specific heats can still be consistent with thermodynamic stability~\cite{Bouchet_2005,KABURAKI199421}. Moreover, a negative specific heat at constant volume indicates that as the system loses energy, its temperature increases, which is a hallmark feature of the black hole thermodynamics~\cite{Wald:1999vt,Carlip:2014pma,Mohd:2023hhu}.
It is important to note that even if we relax the $C_V > 0$ requirement for thermodynamic stability, the key conclusions for the late-time evolution of the $\Lambda$CDM and quintessence models remain unaltered. While there are transient regions in the phase space where the relaxed stability conditions ($C_P > 0$ and $C_V < 0$, yielding $C_P - C_V > 0$) are satisfied, the dynamically stable future attractors lie outside these domains. Consequently, these models still fail to achieve thermodynamic stability at their late-time stable nodes.
 Thus, the status of such thermodynamics interpretation in the cosmological settings still remains an open issue that warrants further investigation.


\section*{Acknowledgments}
We thank Kinjalk Lochan for his helpful comments and suggestions. The research of HSS is supported by the Core Research Grant CRG/2021/003053 from the Science and Engineering Research Board (SERB), India. DM thanks Raman Research Institute for support through postdoctoral fellowship. HSS thanks Raman Research Institute for the hospitality.

\section*{Data availability statement}
The data that support the findings of this study are available from the corresponding author upon reasonable request.
\bibliographystyle{unsrtnat}
\bibliography{Cosmo_Thermo}

@article{Bekenstein:1974ax,
    author = "Bekenstein, Jacob D.",
    title = "{Generalized second law of thermodynamics in black hole physics}",
    doi = "10.1103/PhysRevD.9.3292",
    journal = "Phys. Rev. D",
    volume = "9",
    pages = "3292--3300",
    year = "1974"
}

@article{Duary:2023nnf,
    author = "Duary, Tanima and Banerjee, Narayan and Dasgupta, Ananda",
    title = "{Signature flip in deceleration parameter: a thermodynamic phase transition?}",
    eprint = "2303.14031",
    archivePrefix = "arXiv",
    primaryClass = "gr-qc",
    doi = "10.1140/epjc/s10052-023-11995-w",
    journal = "Eur. Phys. J. C",
    volume = "83",
    number = "9",
    pages = "815",
    year = "2023"
}

@article{Duary:2019dfu,
    author = "Duary, Tanima and Banerjee, Ananda Dasgupta Narayan and Banerjee, Narayan",
    title = "{Thawing and Freezing Quintessence Models: A thermodynamic Consideration}",
    eprint = "1906.10408",
    archivePrefix = "arXiv",
    primaryClass = "gr-qc",
    doi = "10.1140/epjc/s10052-019-7406-z",
    journal = "Eur. Phys. J. C",
    volume = "79",
    number = "11",
    pages = "888",
    year = "2019"
}

@article{Saha:2024mwn,
    author = "Saha, Somnath and Saha, Subhajit and Mahata, Nilanjana",
    title = "{For a flat Universe, $C_P/C_V = -q$ : another coincidence in Cosmology?}",
    eprint = "2409.16328",
    archivePrefix = "arXiv",
    primaryClass = "gr-qc",
    doi = "10.1088/1361-6382/adb2d3",
    journal = "Class. Quant. Grav.",
    volume = "42",
    number = "5",
    pages = "055018",
    year = "2025"
}

@article{Chetry:2019tqd,
    author = "Chetry, Binod and Dutta, Jibitesh and Khyllep, Wompherdeiki",
    title = "{Thermodynamics of scalar field models with kinetic corrections}",
    eprint = "1908.04102",
    archivePrefix = "arXiv",
    primaryClass = "gr-qc",
    doi = "10.1142/S0218271819501633",
    journal = "Int. J. Mod. Phys. D",
    volume = "28",
    number = "15",
    pages = "1950163",
    year = "2019"
}

@article{Barboza:2015bha,
    author = "Barboza, Ed{\'e}sio M. and Nunes, Rafael C. and Abreu, Everton M. C. and Neto, Jorge Ananias",
    title = "{Thermodynamic aspects of dark energy fluids}",
    eprint = "1501.03491",
    archivePrefix = "arXiv",
    primaryClass = "gr-qc",
    doi = "10.1103/PhysRevD.92.083526",
    journal = "Phys. Rev. D",
    volume = "92",
    number = "8",
    pages = "083526",
    year = "2015"
}

@article{Bhandari:2017cow,
    author = "Bhandari, Pritikana and Haldar, Sourav and Chakraborty, Subenoy",
    title = "{Interacting dark energy model and thermal stability}",
    doi = "10.1140/epjc/s10052-017-5417-1",
    journal = "Eur. Phys. J. C",
    volume = "77",
    number = "12",
    pages = "840",
    year = "2017"
}

@article{Kodama:1979vn,
    author = "Kodama, Hideo",
    title = "{Conserved Energy Flux for the Spherically Symmetric System and the Back Reaction Problem in the Black Hole Evaporation}",
    reportNumber = "KUNS-506",
    doi = "10.1143/PTP.63.1217",
    journal = "Prog. Theor. Phys.",
    volume = "63",
    pages = "1217",
    year = "1980"
}

@article{Hayward:1997jp,
    author = "Hayward, Sean A.",
    title = "{Unified first law of black hole dynamics and relativistic thermodynamics}",
    eprint = "gr-qc/9710089",
    archivePrefix = "arXiv",
    doi = "10.1088/0264-9381/15/10/017",
    journal = "Class. Quant. Grav.",
    volume = "15",
    pages = "3147--3162",
    year = "1998"
}

@article{Hayward:2008jq,
    author = "Hayward, S. A. and Di Criscienzo, R. and Vanzo, L. and Nadalini, M. and Zerbini, S.",
    title = "{Local Hawking temperature for dynamical black holes}",
    eprint = "0806.0014",
    archivePrefix = "arXiv",
    primaryClass = "gr-qc",
    doi = "10.1088/0264-9381/26/6/062001",
    journal = "Class. Quant. Grav.",
    volume = "26",
    pages = "062001",
    year = "2009"
}

@article{Bousso:2004tv,
    author = "Bousso, Raphael",
    title = "{Cosmology and the S-matrix}",
    eprint = "hep-th/0412197",
    archivePrefix = "arXiv",
    reportNumber = "UCB-PTH-04-36",
    doi = "10.1103/PhysRevD.71.064024",
    journal = "Phys. Rev. D",
    volume = "71",
    pages = "064024",
    year = "2005"
}

@article{Mimoso:2016jwg,
    author = "Mimoso, Jos{\'e} Pedro and Pav{\'o}n, Diego",
    title = "{Considerations on the thermal equilibrium between matter and the cosmic horizon}",
    eprint = "1610.07788",
    archivePrefix = "arXiv",
    primaryClass = "gr-qc",
    doi = "10.1103/PhysRevD.94.103507",
    journal = "Phys. Rev. D",
    volume = "94",
    number = "10",
    pages = "103507",
    year = "2016"
}

@article{Izquierdo:2005ku,
    author = "Izquierdo, German and Pavon, Diego",
    title = "{Dark energy and the generalized second law}",
    eprint = "astro-ph/0505601",
    archivePrefix = "arXiv",
    doi = "10.1016/j.physletb.2005.12.040",
    journal = "Phys. Lett. B",
    volume = "633",
    pages = "420--426",
    year = "2006"
}

@article{Setare:2006vz,
    author = "Setare, M. R. and Shafei, S.",
    title = "{The Holographic model of dark energy and thermodynamics of non-flat accelerated expanding universe}",
    eprint = "gr-qc/0606103",
    archivePrefix = "arXiv",
    doi = "10.1088/1475-7516/2006/09/011",
    journal = "JCAP",
    volume = "09",
    pages = "011",
    year = "2006"
}

@article{Gong:2006ma,
    author = "Gong, Yungui and Wang, Bin and Wang, Anzhong",
    title = "{Thermodynamical properties of the Universe with dark energy}",
    eprint = "gr-qc/0610151",
    archivePrefix = "arXiv",
    doi = "10.1088/1475-7516/2007/01/024",
    journal = "JCAP",
    volume = "01",
    pages = "024",
    year = "2007"
}

@article{Cai:2005ra,
    author = "Cai, Rong-Gen and Kim, Sang Pyo",
    title = "{First law of thermodynamics and Friedmann equations of Friedmann-Robertson-Walker universe}",
    eprint = "hep-th/0501055",
    archivePrefix = "arXiv",
    doi = "10.1088/1126-6708/2005/02/050",
    journal = "JHEP",
    volume = "02",
    pages = "050",
    year = "2005"
}

@article{Luciano:2022ffn,
    author = "Luciano, Giuseppe Gaetano",
    title = "{Cosmic evolution and thermal stability of Barrow holographic dark energy in a nonflat Friedmann-Robertson-Walker Universe}",
    eprint = "2210.06320",
    archivePrefix = "arXiv",
    primaryClass = "gr-qc",
    doi = "10.1103/PhysRevD.106.083530",
    journal = "Phys. Rev. D",
    volume = "106",
    number = "8",
    pages = "083530",
    year = "2022"
}

@article{Padmanabhan:1989gm,
    author = "Padmanabhan, T.",
    title = "{Statistical Mechanics of Gravitating Systems}",
    reportNumber = "TIFR-TAP-14/89",
    doi = "10.1016/0370-1573(90)90051-3",
    journal = "Phys. Rept.",
    volume = "188",
    pages = "285",
    year = "1990"
}

@article{Chakrabarti:2025qsv,
    author = "Chakrabarti, Soumya",
    title = "{Acceleration from a phase of entropic balance}",
    eprint = "2504.00651",
    archivePrefix = "arXiv",
    primaryClass = "gr-qc",
    doi = "10.1140/epjc/s10052-025-14335-2",
    journal = "Eur. Phys. J. C",
    volume = "85",
    number = "5",
    pages = "585",
    year = "2025"
}

@article{Chakrabarti:2024amb,
    author = "Chakrabarti, Soumya",
    title = "{Phase Transition and Thermodynamic Stability in an Entropy-Driven Universe}",
    eprint = "2403.12622",
    archivePrefix = "arXiv",
    primaryClass = "gr-qc",
    doi = "10.1002/prop.202400063",
    journal = "Fortsch. Phys.",
    volume = "72",
    number = "11",
    pages = "2400063",
    year = "2024"
}

@article{Samaddar_2025,
author = {Samaddar, Amit and Singh, S. Surendra},
title = {Dynamical System Approach and Thermodynamical Perspective of Hořava-Lifshitz Gravity},
journal = {Fortschritte der Physik},
volume = {72},
number = {6},
pages = {2400006},
doi = {https://doi.org/10.1002/prop.202400006},
url = {https://onlinelibrary.wiley.com/doi/abs/10.1002/prop.202400006},
eprint = {https://onlinelibrary.wiley.com/doi/pdf/10.1002/prop.202400006},
year = {2024}
}

@article{Bahamonde:2018pr,
title = {Dynamical systems applied to cosmology: Dark energy and modified gravity},
journal = {Physics Reports},
volume = {775-777},
pages = {1-122},
year = {2018},
note = {Dynamical systems applied to cosmology: Dark energy and modified gravity},
issn = {0370-1573},
doi = {https://doi.org/10.1016/j.physrep.2018.09.001},
url = {https://www.sciencedirect.com/science/article/pii/S0370157318302242},
author = {Sebastian Bahamonde and Christian G. Böhmer and Sante Carloni and Edmund J. Copeland and Wei Fang and Nicola Tamanini},
}

@ARTICLE{Copeland:1998PhRvD,
       author = {{Copeland}, Edmund J. and {Liddle}, Andrew R. and {Wands}, David},
        title = "{Exponential potentials and cosmological scaling solutions}",
      journal = {Phys. Rev. D},
         year = 1998,
        month = apr,
       volume = {57},
       number = {8},
        pages = {4686-4690},
          doi = {10.1103/PhysRevD.57.4686},
archivePrefix = {arXiv},
       eprint = {gr-qc/9711068},
 primaryClass = {gr-qc},
       adsurl = {https://ui.adsabs.harvard.edu/abs/1998PhRvD..57.4686C}
}

@ARTICLE{Scherrer:2008PhRvD,
       author = {{Scherrer}, Robert J. and {Sen}, A.~A.},
        title = "{Thawing quintessence with a nearly flat potential}",
      journal = {Phys. Rev. D},
         year = 2008,
        month = apr,
       volume = {77},
       number = {8},
          eid = {083515},
        pages = {083515},
          doi = {10.1103/PhysRevD.77.083515},
archivePrefix = {arXiv},
       eprint = {0712.3450},
 primaryClass = {astro-ph},
       adsurl = {https://ui.adsabs.harvard.edu/abs/2008PhRvD..77h3515S}
}

@ARTICLE{Gong:2014PhLB,
       author = {{Gong}, Yungui},
        title = "{The general property of dynamical quintessence field}",
      journal = {Physics Letters B},
         year = 2014,
        month = apr,
       volume = {731},
        pages = {342-349},
          doi = {10.1016/j.physletb.2014.03.013},
archivePrefix = {arXiv},
       eprint = {1401.1959},
 primaryClass = {gr-qc},
       adsurl = {https://ui.adsabs.harvard.edu/abs/2014PhLB..731..342G}
}

@ARTICLE{Qi:2016IJTP,
       author = {{Qi}, Jing-Zhao and {Zhang}, Ming-Jian and {Liu}, Wen-Biao},
        title = "{Dynamical Evolution of Quintessence Cosmology in a Physical Phase Space}",
      journal = {International Journal of Theoretical Physics},
         year = 2016,
        month = aug,
       volume = {55},
       number = {8},
        pages = {3672-3681},
          doi = {10.1007/s10773-016-2996-9},
archivePrefix = {arXiv},
       eprint = {1504.04582},
 primaryClass = {gr-qc},
       adsurl = {https://ui.adsabs.harvard.edu/abs/2016IJTP...55.3672Q}
}

@ARTICLE{Steinhardt:1999PhRvD,
       author = {{Steinhardt}, Paul J. and {Wang}, Limin and {Zlatev}, Ivaylo},
        title = "{Cosmological tracking solutions}",
      journal = {Phys. Rev. D},
         year = 1999,
        month = jun,
       volume = {59},
       number = {12},
          eid = {123504},
        pages = {123504},
          doi = {10.1103/PhysRevD.59.123504},
archivePrefix = {arXiv},
       eprint = {astro-ph/9812313},
 primaryClass = {astro-ph},
       adsurl = {https://ui.adsabs.harvard.edu/abs/1999PhRvD..59l3504S}
}

@ARTICLE{Macorra:2000PhRvD,
       author = {{de la Macorra}, A. and {Piccinelli}, G.},
        title = "{Cosmological evolution of general scalar fields and quintessence}",
      journal = {Phys. Rev. D},
         year = 2000,
        month = jun,
       volume = {61},
       number = {12},
          eid = {123503},
        pages = {123503},
          doi = {10.1103/PhysRevD.61.123503},
archivePrefix = {arXiv},
       eprint = {hep-ph/9909459},
 primaryClass = {hep-ph},
       adsurl = {https://ui.adsabs.harvard.edu/abs/2000PhRvD..61l3503D}
}

@ARTICLE{Ng:2001PhRvD,
       author = {{Ng}, S.~C.~C. and {Nunes}, N.~J. and {Rosati}, F.},
        title = "{Applications of scalar attractor solutions to cosmology}",
      journal = {Phys. Rev. D},
         year = 2001,
        month = oct,
       volume = {64},
       number = {8},
        pages = {083510},
          doi = {10.1103/PhysRevD.64.083510},
archivePrefix = {arXiv},
       eprint = {astro-ph/0107321},
 primaryClass = {astro-ph},
       adsurl = {https://ui.adsabs.harvard.edu/abs/2001PhRvD..64h3510N}
}

@ARTICLE{Halliwell:1987PhLB,
       author = {{Halliwell}, J.~J.},
        title = "{Scalar fields in cosmology with an exponential potential}",
      journal = {Physics Letters B},
         year = 1987,
        month = feb,
       volume = {185},
       number = {3-4},
        pages = {341-344},
          doi = {10.1016/0370-2693(87)91011-2},
       adsurl = {https://ui.adsabs.harvard.edu/abs/1987PhLB..185..341H}
}

@ARTICLE{Burd:1988NuPhB,
       author = {{Burd}, A.~B. and {Barrow}, John D.},
        title = "{Inflationary models with exponential potentials}",
      journal = {Nuclear Physics B},
         year = 1988,
        month = oct,
       volume = {308},
       number = {4},
        pages = {929-945},
          doi = {10.1016/0550-3213(88)90135-6},
       adsurl = {https://ui.adsabs.harvard.edu/abs/1988NuPhB.308..929B}
}

@article{garcia2015introduction,
  title={Introduction to the application of dynamical systems theory in the study of the dynamics of cosmological models of dark energy},
  author={Garc{\'\i}a-Salcedo, Ricardo and Gonzalez, Tame and Horta-Rangel, Francisco A and Quiros, Israel and Sanchez-Guzm{\'a}n, Daniel},
  journal={European Journal of Physics},
  volume={36},
  number={2},
  pages={025008},
  year={2015},
  publisher={IOP Publishing}
}

@article{fay2013ten,
  title={Ten scenarios from early radiation to late time acceleration with a minimally coupled dark energy},
  author={Fay, St{\'e}phane},
  journal={Journal of Cosmology and Astroparticle Physics},
  volume={2013},
  number={09},
  pages={023},
  year={2013},
  publisher={IOP Publishing}
}

@article{faraoni2013scalar,
  title={Scalar field cosmology in phase space},
  author={Faraoni, Valerio and Protheroe, Charles S},
  journal={General Relativity and Gravitation},
  volume={45},
  number={1},
  pages={103--123},
  year={2013},
  publisher={Springer}
}

@article{roy2014tracking,
  title={Tracking quintessence: a dynamical systems study},
  author={Roy, Nandan and Banerjee, Narayan},
  journal={General Relativity and Gravitation},
  volume={46},
  number={1},
  pages={1651},
  year={2014},
  publisher={Springer}
}

@book{amendola2010dark,
  title={Dark energy: theory and observations},
  author={Amendola, Luca and Tsujikawa, Shinji},
  year={2010},
  publisher={Cambridge University Press}
}

@article{tsujikawa2013quintessence,
  title={Quintessence: a review},
  author={Tsujikawa, Shinji},
  journal={Classical and Quantum Gravity},
  volume={30},
  number={21},
  pages={214003},
  year={2013},
  publisher={IOP Publishing}
}

@book{Coley:2003mj,
    author = "Coley, A. A.",
    title = "{Dynamical systems and cosmology}",
    doi = "10.1007/978-94-017-0327-7",
    publisher = "Kluwer",
    address = "Dordrecht, Netherlands",
    year = "2003"
}

@article{copeland2006dynamics,
  title={Dynamics of dark energy},
  author={Copeland, Edmund J and Sami, Mohammad and Tsujikawa, Shinji},
  journal={International Journal of Modern Physics D},
  volume={15},
  number={11},
  pages={1753--1935},
  year={2006},
  publisher={World Scientific}
}

@ARTICLE{Urena-Lopez:2012JCAP,
       author = {{Ure{\~n}a-L{\'o}pez}, L. Arturo},
        title = "{Unified description of the dynamics of quintessential scalar fields}",
      journal = {Journal of Cosmology and Astroparticle Physics},
         year = 2012,
        month = mar,
       volume = {2012},
       number = {3},
          eid = {035},
        pages = {035},
          doi = {10.1088/1475-7516/2012/03/035},
archivePrefix = {arXiv},
       eprint = {1108.4712},
 primaryClass = {astro-ph.CO},
       adsurl = {https://ui.adsabs.harvard.edu/abs/2012JCAP...03..035U}
}

@ARTICLE{Tamanini:2014PhRvD,
       author = {{Tamanini}, Nicola},
        title = "{Dynamics of cosmological scalar fields}",
      journal = {Phys. Rev. D},
         year = 2014,
        month = apr,
       volume = {89},
       number = {8},
          eid = {083521},
        pages = {083521},
          doi = {10.1103/PhysRevD.89.083521},
archivePrefix = {arXiv},
       eprint = {1401.6339},
 primaryClass = {gr-qc},
       adsurl = {https://ui.adsabs.harvard.edu/abs/2014PhRvD..89h3521T}
}

@article{Singh:2015rqa,
    author = "Singh, Suprit and Singh, Parminder",
    title = "{It's a dark, dark world: Background evolution of interacting $\phi$CDM models beyond simple exponential potentials}",
    eprint = "1507.01535",
    archivePrefix = "arXiv",
    primaryClass = "astro-ph.CO",
    doi = "10.1088/1475-7516/2016/05/017",
    journal = "JCAP",
    volume = "05",
    pages = "017",
    year = "2016"
}

@article{Bekenstein1973,
  title = {Black Holes and Entropy},
  author = {Bekenstein, Jacob D.},
  journal = {Phys. Rev. D},
  volume = {7},
  issue = {8},
  pages = {2333--2346},
  numpages = {0},
  year = {1973},
  month = {Apr},
  publisher = {American Physical Society},
  doi = {10.1103/PhysRevD.7.2333},
  url = {https://link.aps.org/doi/10.1103/PhysRevD.7.2333}
}

@article{Bardeen1973,
  doi = {10.1007/bf01645742},
  url = {https://doi.org/10.1007/bf01645742},
  year = {1973},
  month = jun,
  publisher = {Springer Science and Business Media {LLC}},
  volume = {31},
  number = {2},
  pages = {161--170},
  author = {J. M. Bardeen and B. Carter and S. W. Hawking},
  title = {The four laws of black hole mechanics},
  journal = {Communications in Mathematical Physics}
}

@article{Hawking1975,
  doi = {10.1007/bf02345020},
  url = {https://doi.org/10.1007/bf02345020},
  year = {1975},
  month = aug,
  publisher = {Springer Science and Business Media {LLC}},
  volume = {43},
  number = {3},
  pages = {199--220},
  author = {S. W. Hawking},
  title = {Particle creation by black holes},
  journal = {Communications In Mathematical Physics}
}

@article{Hawking:1974rv,
  author       = {Hawking, S. W.},
  title        = {Black hole explosions?},
  journal      = {Nature},
  volume       = {248},
  pages        = {30--31},
  year         = {1974},
  doi          = {10.1038/248030a0}
}

@article{Wald:1999vt,
  author    = {Wald, Robert M.},
  title     = {The thermodynamics of black holes},
  journal   = {Living Rev. Rel.},
  volume    = {4},
  pages     = {6},
  year      = {2001},
  eprint    = {gr-qc/9912119},
  doi       = {10.12942/lrr-2001-6}
}

@article{Carlip:2014pma,
  author    = {Carlip, Steven},
  title     = {Black Hole Thermodynamics},
  journal   = {Int. J. Mod. Phys. D},
  volume    = {23},
  pages     = {1430023},
  year      = {2014},
  eprint    = {1410.1486},
  archivePrefix = {arXiv},
  primaryClass = {gr-qc},
  doi       = {10.1142/S0218271814300237}
}

@article{Gibbons:1977mu,
  author    = {Gibbons, G. W. and Hawking, S. W.},
  title     = {Cosmological Event Horizons, Thermodynamics, and Particle Creation},
  journal   = {Phys. Rev. D},
  volume    = {15},
  pages     = {2738--2751},
  year      = {1977},
  doi       = {10.1103/PhysRevD.15.2738}
}

@article{Unruh:1976db,
  author    = {Unruh, W. G.},
  title     = {Notes on black-hole evaporation},
  journal   = {Phys. Rev. D},
  volume    = {14},
  pages     = {870--892},
  year      = {1976},
  doi       = {10.1103/PhysRevD.14.870}
}

@article{Padmanabhan:2009vy,
  author    = {Padmanabhan, T.},
  title     = {Thermodynamical Aspects of Gravity: New insights},
  journal   = {Rept. Prog. Phys.},
  volume    = {73},
  pages     = {046901},
  year      = {2010},
  eprint    = {0911.5004},
  archivePrefix = {arXiv},
  primaryClass = {gr-qc},
  doi       = {10.1088/0034-4885/73/4/046901}
}

@article{Padmanabhan:2022jjt,
  author    = {Padmanabhan, T. and Parattu, Krishnamohan},
  title     = {Gravity and the Thermodynamics of Horizons},
  journal   = {Gen. Rel. Grav.},
  volume    = {54},
  number    = {10},
  pages     = {133},
  year      = {2022},
  eprint    = {2205.05862},
  archivePrefix = {arXiv},
  primaryClass = {gr-qc},
  doi       = {10.1007/s10714-022-02978-y}
}

@article{Mohd:2023hhu,
  author    = {Mohd, Arif and Sarkar, Sudipta},
  title     = {Black Hole Thermodynamics: A Review},
  journal   = {Universe},
  volume    = {9},
  number    = {11},
  pages     = {501},
  year      = {2023},
  eprint    = {2311.02534},
  archivePrefix = {arXiv},
  primaryClass = {gr-qc},
  doi       = {10.3390/universe9110501}
}

@book{Wald:1984rg,
  author    = {Wald, Robert M.},
  title     = {General Relativity},
  publisher = {University of Chicago Press},
  year      = {1984},
  isbn      = {978-0-226-87033-5},
  address   = {Chicago, USA}
}

@ARTICLE{Planck:2020AA,
       author = {{Planck Collaboration}},
        title = "{Planck 2018 results. VI. Cosmological parameters}",
      journal = {Astronomy and Astrophysics},
         year = 2020,
        month = sep,
       volume = {641},
          eid = {A6},
        pages = {A6},
          doi = {10.1051/0004-6361/201833910},
archivePrefix = {arXiv},
       eprint = {1807.06209},
 primaryClass = {astro-ph.CO},
       adsurl = {https://ui.adsabs.harvard.edu/abs/2020A&A...641A...6P}
}

@book{Callen:1985,
  author    = {Callen, Herbert B.},
  title     = {Thermodynamics and an Introduction to Thermostatistics},
  edition   = {2nd},
  publisher = {Wiley},
  year      = {1985},
  isbn      = {978-0-471-86256-7},
  address   = {New York, USA}
}

@book{Blundell:2010,
  author    = {Blundell, Stephen J. and Blundell, Katherine M.},
  title     = {Concepts in Thermal Physics},
  edition   = {2nd},
  publisher = {Oxford University Press},
  year      = {2010},
  isbn      = {978-0-19-956209-1},
  address   = {Oxford, UK}
}

@article{Kong:2021dqd,
    author = "Kong, Shi-Bei and Abdusattar, Haximjan and Yin, Yihao and Zhang, Hongsheng and Hu, Ya-Peng",
    title = "{The $P-V$ phase transition of the FRW universe}",
    eprint = "2108.09411",
    archivePrefix = "arXiv",
    primaryClass = "gr-qc",
    doi = "10.1140/epjc/s10052-022-10976-9",
    journal = "Eur. Phys. J. C",
    volume = "82",
    number = "11",
    pages = "1047",
    year = "2022"
}

@article{Abdusattar:2021wfv,
    author = "Abdusattar, Haximjan and Kong, Shi-Bei and You, Wen-Long and Zhang, Hongsheng and Hu, Ya-Peng",
    title = "{First principle study of gravitational pressure and thermodynamics of FRW universe}",
    eprint = "2108.09407",
    archivePrefix = "arXiv",
    primaryClass = "gr-qc",
    doi = "10.1007/JHEP12(2022)168",
    journal = "JHEP",
    volume = "12",
    pages = "168",
    year = "2022"
}

@article{Kong:2022xny,
    author = "Kong, Shi-Bei and Abdusattar, Haximjan and Zhang, Hongsheng and Hu, Ya-Peng",
    title = "{Equation of state and Joule-Thomson expansion for the FRW universe in the brane world scenario}",
    eprint = "2208.12603",
    archivePrefix = "arXiv",
    primaryClass = "gr-qc",
    doi = "10.1016/j.nuclphysb.2023.116091",
    journal = "Nucl. Phys. B",
    volume = "987",
    pages = "116091",
    year = "2023"
}

@article{Abdusattar:2023hlj,
    author = "Abdusattar, Haximjan and Kong, Shi-Bei and Zhang, Hongsheng and Hu, Ya-Peng",
    title = "{Phase transitions and critical phenomena for the FRW universe in an effective scalar-tensor theory}",
    eprint = "2301.01938",
    archivePrefix = "arXiv",
    primaryClass = "gr-qc",
    doi = "10.1016/j.dark.2023.101330",
    journal = "Phys. Dark Univ.",
    volume = "42",
    pages = "101330",
    year = "2023"
}

@article{banihashemi2023minus,
  title={The minus sign in the first law of de Sitter horizons},
  author={Banihashemi, Batoul and Jacobson, Ted and Svesko, Andrew and Visser, Manus},
  journal={Journal of High Energy Physics},
  volume={2023},
  number={1},
  pages={1--29},
  year={2023},
  publisher={Springer}
}

@article{Luongo:2012dv,
    author = "Luongo, Orlando and Quevedo, Hernando",
    title = "{Cosmographic study of the universe's specific heat: A landscape for Cosmology?}",
    eprint = "1211.0626",
    archivePrefix = "arXiv",
    primaryClass = "gr-qc",
    doi = "10.1007/s10714-013-1649-z",
    journal = "Gen. Rel. Grav.",
    volume = "46",
    pages = "1649",
    year = "2014"
}

@incollection{dodelson2021concordance,
 author = {Dodelson, Scott and Schmidt, Fabian},
 title = {The concordance model of cosmology},
 publisher = {Elsevier},
 year = {2021},
 pages = {1--19},
 doi = {10.1016/b978-0-12-815948-4.00007-3},
 booktitle = {Modern Cosmology}
}

@article{Banerjee:2016nse,
    author = "Banerjee, Rabin and Majhi, Bibhas Ranjan and Samanta, Saurav",
    title = "{Thermogeometric phase transition in a unified framework}",
    eprint = "1611.06701",
    archivePrefix = "arXiv",
    primaryClass = "gr-qc",
    doi = "10.1016/j.physletb.2017.01.040",
    journal = "Phys. Lett. B",
    volume = "767",
    pages = "25--28",
    year = "2017"
}

@article{Davies:1977bgr,
    author = "Davies, P. C. W.",
    title = "{Thermodynamics of Black Holes}",
    reportNumber = "PRINT-78-0644 (KING'S-COLL.)",
    doi = "10.1098/rspa.1977.0047",
    journal = "Proc. Roy. Soc. Lond. A",
    volume = "353",
    pages = "499--521",
    year = "1977"
}

@article{Davies:1989ey,
    author = "Davies, P. C. W.",
    title = "{Thermodynamic Phase Transitions of {Kerr-Newman} Black Holes in De Sitter Space}",
    reportNumber = "NCL-89-TP9",
    doi = "10.1088/0264-9381/6/12/018",
    journal = "Class. Quant. Grav.",
    volume = "6",
    pages = "1909",
    year = "1989"
}

@article{Shen:2005nu,
    author = "Shen, Jian-yong and Cai, Rong-Gen and Wang, Bin and Su, Ru-Keng",
    title = "{Thermodynamic geometry and critical behavior of black holes}",
    eprint = "gr-qc/0512035",
    archivePrefix = "arXiv",
    doi = "10.1142/S0217751X07034064",
    journal = "Int. J. Mod. Phys. A",
    volume = "22",
    pages = "11--27",
    year = "2007"
}

@article{Czinner:2015eyk,
    author = "Czinner, Viktor G. and Iguchi, Hideo",
    title = "{R{\'e}nyi Entropy and the Thermodynamic Stability of Black Holes}",
    eprint = "1511.06963",
    archivePrefix = "arXiv",
    primaryClass = "gr-qc",
    doi = "10.1016/j.physletb.2015.11.061",
    journal = "Phys. Lett. B",
    volume = "752",
    pages = "306--310",
    year = "2016"
}

@article{Banerjee:2010da,
    author = "Banerjee, Rabin and Ghosh, Sumit and Roychowdhury, Dibakar",
    title = {{New type of phase transition in Reissner Nordstr{\"o}m{\textendash}AdS black hole and its thermodynamic geometry}},
    eprint = "1008.2644",
    archivePrefix = "arXiv",
    primaryClass = "gr-qc",
    doi = "10.1016/j.physletb.2010.12.010",
    journal = "Phys. Lett. B",
    volume = "696",
    pages = "156--162",
    year = "2011"
}

@article{Banerjee:2011cz,
    author = "Banerjee, Rabin and Roychowdhury, Dibakar",
    title = "{Critical phenomena in Born-Infeld AdS black holes}",
    eprint = "1111.0147",
    archivePrefix = "arXiv",
    primaryClass = "gr-qc",
    doi = "10.1103/PhysRevD.85.044040",
    journal = "Phys. Rev. D",
    volume = "85",
    pages = "044040",
    year = "2012"
}

@article{Majhi:2016txt,
    author = "Majhi, Bibhas Ranjan and Samanta, Saurav",
    title = "{P-V criticality of AdS black holes in a general framework}",
    eprint = "1609.06224",
    archivePrefix = "arXiv",
    primaryClass = "gr-qc",
    doi = "10.1016/j.physletb.2017.08.038",
    journal = "Phys. Lett. B",
    volume = "773",
    pages = "203--207",
    year = "2017"
}

@article{Bhattacharya:2019awq,
    author = "Bhattacharya, Krishnakanta and Dey, Sumit and Majhi, Bibhas Ranjan and Samanta, Saurav",
    title = "{General framework to study the extremal phase transition of black holes}",
    eprint = "1903.03434",
    archivePrefix = "arXiv",
    primaryClass = "gr-qc",
    doi = "10.1103/PhysRevD.99.124047",
    journal = "Phys. Rev. D",
    volume = "99",
    number = "12",
    pages = "124047",
    year = "2019"
}

@article{Thomas:2012zzc,
    author = "Thomas, Bouetou Bouetou and Saleh, Mahamat and Kofane, Timoleon Crepin",
    title = "{Thermodynamics and phase transition of the Reissner-Nordstroem black hole surrounded by quintessence}",
    eprint = "1604.06207",
    archivePrefix = "arXiv",
    primaryClass = "gr-qc",
    doi = "10.1007/s10714-012-1382-z",
    journal = "Gen. Rel. Grav.",
    volume = "44",
    pages = "2181--2189",
    year = "2012"
}

@article{Pavon:1991kh,
    author = "Pavon, D.",
    title = "{Phase transition in Reissner-Nordstrom black holes}",
    doi = "10.1103/PhysRevD.43.2495",
    journal = "Phys. Rev. D",
    volume = "43",
    pages = "2495--2497",
    year = "1991"
}

@article{Kaburaki:1993ah,
    author = "Kaburaki, O. and Okamoto, I. and Katz, J.",
    title = "{Thermodynamic Stability of Kerr Black holes}",
    doi = "10.1103/PhysRevD.47.2234",
    journal = "Phys. Rev. D",
    volume = "47",
    pages = "2234--2241",
    year = "1993"
}

@article{Kothawala:2007em,
    author = "Kothawala, Dawood and Sarkar, Sudipta and Padmanabhan, T.",
    title = "{Einstein's equations as a thermodynamic identity: The cases of stationary axisymmetric horizons and evolving spherically symmetric horizons}",
    eprint = "gr-qc/0701002",
    archivePrefix = "arXiv",
    doi = "10.1016/j.physletb.2007.07.021",
    journal = "Phys. Lett. B",
    volume = "652",
    pages = "338--342",
    year = "2007"
}

@article{Katz:1993up,
    author = "Katz, Joseph and Okamoto, I. and Kaburaki, O.",
    title = "{Thermodynamic stability of pure black holes}",
    doi = "10.1088/0264-9381/10/7/009",
    journal = "Class. Quant. Grav.",
    volume = "10",
    pages = "1323--1339",
    year = "1993"
}

@article{Padmanabhan:2002ha,
    author = "Padmanabhan, T.",
    editor = "Ahluwalia, Dharam Vir and Dadhich, N. K.",
    title = "{Thermodynamics and / of horizons: A Comparison of Schwarzschild, Rindler and de Sitter space-times}",
    eprint = "gr-qc/0202078",
    archivePrefix = "arXiv",
    reportNumber = "IUCAA-05-2002, IUCAA-PREPRINT-05-2002",
    doi = "10.1142/S021773230200751X",
    journal = "Mod. Phys. Lett. A",
    volume = "17",
    pages = "923--942",
    year = "2002"
}

@article{Bargueno:2024mys,
    author = "Bargue{\~n}o, Pedro and Fern{\'a}ndez-Silvestre, Diego and Miralles, Juan A.",
    title = "{Physical reinterpretation of heat capacity discontinuities for static black holes}",
    eprint = "2407.10885",
    archivePrefix = "arXiv",
    primaryClass = "gr-qc",
    doi = "10.1103/PhysRevD.110.124035",
    journal = "Phys. Rev. D",
    volume = "110",
    number = "12",
    pages = "124035",
    year = "2024"
}

@article{Sokolowski:1980uva,
    author = "Sokolowski, L. M. and Mazur, P.",
    title = "{Second-order phase transitions in black-hole thermodynamics}",
    doi = "10.1088/0305-4470/13/3/043",
    journal = "J. Phys.",
    volume = "13",
    pages = "A1113--1120",
    year = "1980"
}

@article{Davies:1978zz,
    author = "Davies, P. C. W.",
    title = "{Thermodynamics of black holes}",
    doi = "10.1088/0034-4885/41/8/004",
    journal = "Rept. Prog. Phys.",
    volume = "41",
    pages = "1313--1355",
    year = "1978"
}

@article{Candelas:1977zz,
    author = "Candelas, P. and Sciama, D. W.",
    title = "{Irreversible Thermodynamics of Black Holes}",
    doi = "10.1103/PhysRevLett.38.1372",
    journal = "Phys. Rev. Lett.",
    volume = "38",
    pages = "1372--1375",
    year = "1977"
}

@article{KABURAKI199421,
title = {Should entropy be concave?},
journal = {Physics Letters A},
volume = {185},
number = {1},
pages = {21-28},
year = {1994},
issn = {0375-9601},
doi = {https://doi.org/10.1016/0375-9601(94)90981-4},
url = {https://www.sciencedirect.com/science/article/pii/0375960194909814},
author = {Osamu Kaburaki}
}

@article{HERTEL1971520,
title = {A soluble model for a system with negative specific heat},
journal = {Annals of Physics},
volume = {63},
number = {2},
pages = {520-533},
year = {1971},
issn = {0003-4916},
doi = {https://doi.org/10.1016/0003-4916(71)90025-X},
url = {https://www.sciencedirect.com/science/article/pii/000349167190025X},
author = {P Hertel and W Thirring}
}

@article{Bouchet_2005,
   title={Classification of Phase Transitions and Ensemble Inequivalence, in Systems with Long Range Interactions},
   volume={118},
   ISSN={1572-9613},
   url={http://dx.doi.org/10.1007/s10955-004-2059-0},
   DOI={10.1007/s10955-004-2059-0},
   number={5–6},
   journal={Journal of Statistical Physics},
   publisher={Springer Science and Business Media LLC},
   author={Bouchet, F. and Barr\'e, J.},
   year={2005},
   month=mar, pages={1073–1105} }

@ARTICLE{Shahzad2024-qz,
  title     = "Phase transitions and thermal properties of the charged Acoustic
               black hole",
  author    = "Shahzad, M R and Abbas, G and Zhu, Tao and Ali, R H and Ashraf,
               Asifa",
  journal   = "Phys. Dark Universe",
  publisher = "Elsevier BV",
  volume    =  46,
  number    =  101694,
  pages     = "101694",
  month     =  dec,
  year      =  2024,
  language  = "en"
}

@ARTICLE{Panda2024-dg,
  title     = "Thermodynamics of a non-canonical {f(R,T)} gravity",
  author    = "Panda, Arijit and Manna, Goutam and Ray, Saibal and Khlopov,
               Maxim and Dhankar, Praveen Kumar",
  journal   = "Phys. Dark Universe",
  publisher = "Elsevier BV",
  volume    =  46,
  number    =  101697,
  pages     = "101697",
  month     =  dec,
  year      =  2024,
  language  = "en"
}

@ARTICLE{Singh2024-wq,
  title     = "Power law cosmology in modified theory with thermodynamics
               analysis",
  author    = "Singh, J K and {Shaily} and Pradhan, Anirudh and Beesham,
               Aroonkumar",
  journal   = "Phys. Dark Universe",
  publisher = "Elsevier BV",
  volume    =  46,
  number    =  101658,
  pages     = "101658",
  month     =  dec,
  year      =  2024,
  language  = "en"
}

@ARTICLE{Cruz2024-iq,
  title     = "A new approach to {P-V} phase transitions: Einstein gravity and
               holographic type dark energy",
  author    = "Cruz, Miguel and Lepe, Samuel and Saavedra, Joel",
  journal   = "Phys. Dark Universe",
  publisher = "Elsevier BV",
  volume    =  46,
  number    =  101580,
  pages     = "101580",
  month     =  dec,
  year      =  2024,
  language  = "en"
}

@ARTICLE{Cruz2023-sf,
  title     = "Thermodynamics of a transient phantom scenario",
  author    = "Cruz, Miguel and Lepe, Samuel",
  journal   = "Phys. Dark Universe",
  publisher = "Elsevier BV",
  volume    =  42,
  number    =  101367,
  pages     = "101367",
  month     =  dec,
  year      =  2023,
  language  = "en"
}

@ARTICLE{Sebastiani2023-sr,
  title     = "First law of thermodynamics and entropy of {FLRW} universe in
               modified gravity",
  author    = "Sebastiani, Lorenzo",
  journal   = "Phys. Dark Universe",
  publisher = "Elsevier BV",
  volume    =  42,
  number    =  101296,
  pages     = "101296",
  month     =  dec,
  year      =  2023,
  language  = "en"
}

@ARTICLE{Bahamonde2018-qw,
  title     = "Thermodynamics and cosmological reconstruction in {f(T,B)} gravity",
  author    = "Bahamonde, Sebastian and Zubair, M and Abbas, G",
  journal   = "Phys. Dark Universe",
  publisher = "Elsevier BV",
  volume    =  19,
  pages     = "78--90",
  month     =  mar,
  year      =  2018
}

@article{Caro:2025,
author = {Rivadeneira-Caro, Rodrigo and Saavedra, Joel F. and Tello-Ortiz, Francisco},
title = {Cosmological FLRW Phase Transitions Under Exponential Corrected Entropy},
journal = {Fortschritte der Physik},
volume = {n/a},
number = {n/a},
pages = {e70063},
year = 2025,
doi = {https://doi.org/10.1002/prop.70063},
url = {https://onlinelibrary.wiley.com/doi/abs/10.1002/prop.70063},
eprint = {https://onlinelibrary.wiley.com/doi/pdf/10.1002/prop.70063}
}

@article{Caldwell:1999ew,
    author = "Caldwell, R. R.",
    title = "{A Phantom menace?}",
    eprint = "astro-ph/9908168",
    archivePrefix = "arXiv",
    doi = "10.1016/S0370-2693(02)02589-3",
    journal = "Phys. Lett. B",
    volume = "545",
    pages = "23--29",
    year = "2002"
}

@article{Caldwell:2003vq,
    author = "Caldwell, Robert R. and Kamionkowski, Marc and Weinberg, Nevin N.",
    title = "{Phantom energy and cosmic doomsday}",
    eprint = "astro-ph/0302506",
    archivePrefix = "arXiv",
    doi = "10.1103/PhysRevLett.91.071301",
    journal = "Phys. Rev. Lett.",
    volume = "91",
    pages = "071301",
    year = "2003"
}

@article{Carroll:2003st,
    author = "Carroll, Sean M. and Hoffman, Mark and Trodden, Mark",
    title = "{Can the dark energy equation-of-state parameter $w$ be less than -1?}",
    eprint = "astro-ph/0301273",
    archivePrefix = "arXiv",
    reportNumber = "EFI-2003-01, SU-GP-03-1-1",
    doi = "10.1103/PhysRevD.68.023509",
    journal = "Phys. Rev. D",
    volume = "68",
    pages = "023509",
    year = "2003"
}

@article{Cline:2003gs,
    author = "Cline, James M. and Jeon, Sangyong and Moore, Guy D.",
    title = "{The Phantom menaced: Constraints on low-energy effective ghosts}",
    eprint = "hep-ph/0311312",
    archivePrefix = "arXiv",
    reportNumber = "MCGILL-03-25",
    doi = "10.1103/PhysRevD.70.043543",
    journal = "Phys. Rev. D",
    volume = "70",
    pages = "043543",
    year = "2004"
}

@ARTICLE{Valentino:2020JCAP,
       author = {{Di Valentino}, Eleonora and {Melchiorri}, Alessandro and {Silk}, Joseph},
        title = "{Cosmological constraints in extended parameter space from the Planck 2018 Legacy release}",
      journal = {Journal of Cosmology and Astroparticle Physics},
         year = 2020,
        month = jan,
       volume = {2020},
       number = {1},
          eid = {013},
        pages = {013},
          doi = {10.1088/1475-7516/2020/01/013},
archivePrefix = {arXiv},
       eprint = {1908.01391},
 primaryClass = {astro-ph.CO},
       adsurl = {https://ui.adsabs.harvard.edu/abs/2020JCAP...01..013D}
}

@ARTICLE{DESI:2025JCAP,
       author = {DESI Collaboration},
        title = "{DESI 2024 VI: cosmological constraints from the measurements of baryon acoustic oscillations}",
      journal = {Journal of Cosmology and Astroparticle Physics},
         year = 2025,
        month = feb,
       volume = {2025},
       number = {2},
          eid = {021},
        pages = {021},
          doi = {10.1088/1475-7516/2025/02/021},
archivePrefix = {arXiv},
       eprint = {2404.03002},
 primaryClass = {astro-ph.CO},
       adsurl = {https://ui.adsabs.harvard.edu/abs/2025JCAP...02..021A}
}

@ARTICLE{Chakraborty:2025JCAP,
       author = {{Chakraborty}, Amlan and {Chanda}, Prolay and {Das}, Subinoy and {Dutta}, Koushik},
        title = "{DESI results: hint towards coupled dark matter and dark energy}",
      journal = {Journal of Cosmology and Astroparticle Physics},
         year = 2025,
        month = nov,
       volume = {2025},
       number = {11},
          eid = {047},
        pages = {047},
          doi = {10.1088/1475-7516/2025/11/047},
archivePrefix = {arXiv},
       eprint = {2503.10806},
 primaryClass = {astro-ph.CO},
       adsurl = {https://ui.adsabs.harvard.edu/abs/2025JCAP...11..047C}
}

@article{Basilakos:2019dof,
    author = "Basilakos, Spyros and Leon, Genly and Papagiannopoulos, G. and Saridakis, Emmanuel N.",
    title = "{Dynamical system analysis at background and perturbation levels: Quintessence in severe disadvantage comparing to $\Lambda$CDM}",
    eprint = "1904.01563",
    archivePrefix = "arXiv",
    primaryClass = "gr-qc",
    doi = "10.1103/PhysRevD.100.043524",
    journal = "Phys. Rev. D",
    volume = "100",
    number = "4",
    pages = "043524",
    year = "2019"
}

@article{Jamil:2009eb,
    author = "Jamil, Mubasher and Saridakis, Emmanuel N. and Setare, M. R.",
    title = "{Thermodynamics of dark energy interacting with dark matter and radiation}",
    eprint = "0910.0822",
    archivePrefix = "arXiv",
    primaryClass = "hep-th",
    doi = "10.1103/PhysRevD.81.023007",
    journal = "Phys. Rev. D",
    volume = "81",
    pages = "023007",
    year = "2010"
}

@article{Xu:2012jf,
    author = "Xu, Chen and Saridakis, Emmanuel N. and Leon, Genly",
    title = "{Phase-Space analysis of Teleparallel Dark Energy}",
    eprint = "1202.3781",
    archivePrefix = "arXiv",
    primaryClass = "gr-qc",
    doi = "10.1088/1475-7516/2012/07/005",
    journal = "JCAP",
    volume = "07",
    pages = "005",
    year = "2012"
}

@article{Leon:2012mt,
    author = "Leon, Genly and Saridakis, Emmanuel N.",
    title = "{Dynamical analysis of generalized Galileon cosmology}",
    eprint = "1211.3088",
    archivePrefix = "arXiv",
    primaryClass = "astro-ph.CO",
    doi = "10.1088/1475-7516/2013/03/025",
    journal = "JCAP",
    volume = "03",
    pages = "025",
    year = "2013"
}

@article{Leon:2013qh,
    author = "Leon, Genly and Saavedra, Joel and Saridakis, Emmanuel N.",
    title = "{Cosmological behavior in extended nonlinear massive gravity}",
    eprint = "1301.7419",
    archivePrefix = "arXiv",
    primaryClass = "astro-ph.CO",
    reportNumber = "CLASS.QUANT.GRAV.-30-(2013)-135001",
    doi = "10.1088/0264-9381/30/13/135001",
    journal = "Class. Quant. Grav.",
    volume = "30",
    pages = "135001",
    year = "2013"
}

@article{Leon:2014yua,
    author = "Leon, Genly and Saridakis, Emmanuel N.",
    title = "{Dynamical behavior in mimetic F(R) gravity}",
    eprint = "1501.00488",
    archivePrefix = "arXiv",
    primaryClass = "gr-qc",
    doi = "10.1088/1475-7516/2015/04/031",
    journal = "JCAP",
    volume = "04",
    pages = "031",
    year = "2015"
}

\end{document}